\documentclass[prd,notitlepage,showpacs,preprintnumbers,amsmath,amssymb,nofootinbib,APS,11pt,superscriptaddress]{revtex4-2}

\usepackage{dcolumn}
\usepackage{bm}
\usepackage{xcolor}
\usepackage[utf8]{inputenc}
\usepackage[spanish,english]{babel}
\usepackage{amsmath,amssymb,amsfonts,latexsym,cancel}
\usepackage{graphicx}
\usepackage{color}
\usepackage{soul}
\usepackage{ulem}
\usepackage{hyperref}
\usepackage{amsmath}
\usepackage{slashed}
\usepackage{braket}
\usepackage{amssymb}
\usepackage{amsmath}
\allowdisplaybreaks 


\newcommand{\be}{\begin{equation}}
\newcommand{\ee}{\end{equation}}
\newcommand{\bea}{\begin{eqnarray}}
\newcommand{\eea}{\end{eqnarray}}

\newcommand{\langleRE}{\langle{:}\,}
\newcommand{\rangleRE}{\,{:}\rangle}

\newcommand{\SprPhi}{:\hspace{-0.1cm}\Delta_{\phi^2}\hspace{-0.1cm}:}
\newcommand{\SprRho}{:\hspace{-0.1cm}\Delta_{\rho}\hspace{-0.1cm}:}
\newcommand{\SprPre}{:\hspace{-0.1cm}\Delta_{p}\hspace{-0.1cm}:}
\newcommand{\SprGen}{:\hspace{-0.1cm}\Delta_{c}\hspace{-0.1cm}:}

\newcommand{\HOne}{\hspace{-0.05cm} ^{^{(1)}\hspace{-0.1cm}}H_{ab}}
\newcommand{\HTwo}{\hspace{-0.05cm} ^{^{(2)}\hspace{-0.1cm}}H_{ab}}

\newcommand{\omegaa}{w_{\mathit{2}}}
\newcommand{\omegab}{w_{\mathit{4}}}

\newcommand{\Mma}{M_{\mathit{2}}}
\newcommand{\Mmb}{M_{\mathit{4}}}

\newcommand{\MmaH}{M_{\mathit{2}H}}
\newcommand{\MmbH}{M_{\mathit{4}H}}

\begin{document}

\title{\bf \Large Physical scale adiabatic regularization in cosmological spacetimes}

\author{Antonio Ferreiro}\email{antonio.ferreiro@ru.nl}
\affiliation{Freudenthal Institute. Utrecht University, 3584CC Utrecht, The Netherlands}
\author{Samuel Monin}\email{samumonin@icloud.com}
\affiliation{Departamento de Fisica Teórica and IFIC (Universidad de Valencia-CSIC). Universidad de Valencia, Burjassot-46100, Valencia, Spain}
\author{Francisco Torrenti}\email{f.torrenti@uv.es}
\affiliation{Departamento de Fisica Teórica and IFIC (Universidad de Valencia-CSIC). Universidad de Valencia, Burjassot-46100, Valencia, Spain}

\begin{abstract}
We develop a new regularization method for the stress-energy tensor and the two-point function of free quantum scalar fields propagating in cosmological spacetimes. We proceed by extending the adiabatic regularization scheme with the introduction of two additional mass scales. By setting them to the order of the physical scale of the studied scenario, we obtain ultraviolet-regularized quantities that do not distort the power spectra amplitude at the infrared scales amplified by the expansion of the universe. This is not ensured by the standard adiabatic approach. We also show how our proposed subtraction terms can be interpreted as a renormalization of coupling constants in the Einstein equations. We finally illustrate our proposed regularization method in two scenarios of cosmological interest: de Sitter inflation and geometric reheating.
\end{abstract}

\date{\today}
\maketitle

\tableofcontents\newpage

\section{Introduction}

Quantum field theory in curved spacetime provides the most adequate framework to study the dynamics of quantum fields propagating on classical background spacetimes \cite{Parker:2009uva,Wald:1995yp,Birrell:1982ix,Hu:2020luk}. A fundamental implication of the theory is the one of gravitational particle production, which takes place e.g.~when the Universe expands non-adiabatically \cite{Parker:1968mv,Parker:1969au,Parker:1971pt} or near the event horizon of a black hole \cite{Hawking:1975vcx}. In the cosmological context, this provides a natural mechanism for the excitation of metric fluctuations during inflation \cite{Guth:1980zm,Linde:1981mu,Albrecht:1982wi,Starobinsky:1980te}, which constitute the seeds for the formation of structure in the universe \cite{Sasaki:1986hm,Mukhanov:1988jd}. The excited tensor perturbations also constitute a relevant source of primordial gravitational waves, whose amplitude is observationally upper bounded by CMB experiments \cite{BICEP:2021xfz}. Gravitational effects can also play an important role in the production of particles at the end of inflation \cite{Ford:1986sy} or during reheating \cite{Bassett:1997az}. For a recent review on cosmological particle production, see \cite{Ford:2021syk}.

An essential problem when working with quantum fields in curved spacetimes is the one of renormalization, which is challenging even for free fields.  For instance, if we consider a free scalar field $\phi$ and compute the expectation value of its stress-energy tensor $\langle T_{\mu \nu}(x) \rangle $, we find new ultraviolet divergences that are not present in flat spacetime, and hence cannot be removed by a normal ordering procedure. A similar problem appears in the two-point function $\langle \phi (x) \phi (x') \rangle$ when evaluated at coincident spacetime points $x \to x'$. Several regularization and renormalization methods to deal with these divergences have been developed for quantum fields in curved spacetimes, see e.g.~\cite{Parker:2009uva,Birrell:1982ix} for standard textbooks on the subject. 

In this work we will focus on regularization in cosmological spacetimes, described by Friedmann-Lemaitre-Robertson-Walker (FLRW) metrics. An extensively used regularization scheme in these spacetimes is \textit{adiabatic regularization} \cite{Parker:1974qw,Fulling:1974zr,Fulling:1974pu}, which is based on an adiabatic WKB-like expansion of the field modes. Given an unregularized quantity, one can identify its ultraviolet-divergent contributions by expanding it adiabatically up to a certain order, and then subtract the obtained terms to obtain a finite expression. Despite being specific to FLRW metrics, the method is equivalent to general curved background constructions \cite{delRio:2014bpa, Beltran-Palau:2021xrk}, which ensures that observables are constructed in a local covariant way. It is also very convenient to use in numerical computations (see e.g.~\cite{Anderson:1987yt,Habib:1999cs,Zago:2018huk}) and can also be applied to the construction of preferred vacuum states \cite{Agullo:2014ica, Ferreiro:2022hik}. The method has been extended to the regularization of spin-1/2 fields \cite{Landete:2013axa, Landete:2013lpa, delRio:2014cha, BarberoG:2018oqi} and spin-1 fields \cite{Maranon-Gonzalez:2023efu}.

However, although the subtraction terms obtained through the adiabatic expansion successfully remove the ultraviolet divergences, it has been shown that they can distort the amplitude at infrared scales, especially in the case of light scalar fields. For example, the regularized power spectrum of a light field in de Sitter spacetime ($m\ll H$) gets significantly suppressed at scales  $k \gtrsim a m$, and it is exactly zero in the massless limit $m \rightarrow 0$ \cite{Parker:2007ni}. These results can potentially change the standard observable predictions of slow-roll inflation \cite{Agullo:2008ka, Agullo:2009vq,Agullo:2009zi}, and have been critically examined in several works \cite{Finelli:2007fr,Durrer:2009ii,Urakawa:2009xaa,Marozzi:2011da,Agullo:2011qg,Bastero-Gil:2013nja} (see also \cite{Woodard:2014jba}). This is still considered an open problem, as reflected by more recent studies that have tackled this issue from different perspectives. For example, in \cite{Markkanen:2017rvi} it was emphasized that different renormalizations can be found by expressing the subtraction terms in de Sitter space as counterterms in the Lagrangian. In \cite{Animali:2022lig}, an infrared cutoff was introduced in the adiabatic subtraction terms to tame the infrared distortions, and in \cite{Corba:2022ugu} the authors developed a different method based on a resummation of the entire adiabatic expansion.

In any case, it is important to emphasize that the regularization program is in principle ambiguous, and different methods can indeed produce different results. This is already the case of perturbative quantum field theory in Minkowski spacetime, where this ambiguity is encapsulated in the renormalized coupling constants (we will later come back to this point in the case of the gravitational coupling constants). Motivated by this idea, in Ref.~\cite{Ferreiro:2022ibf} we developed a new regularization method for the two-point function at coincident spacetime points, based on an extension of the standard adiabatic scheme. We proposed a new set of subtraction terms that successfully cancel the ultraviolet divergences but also minimize the introduced distortions at momenta scales $k \lesssim M$, where $M$ is an arbitrary mass scale fixed by our choice of regularization scheme. This new method is compatible with the ambiguities allowed by the renormalization program and is consistent with local covariance (unlike e.g.~introducing a hard infrared cutoff in the adiabatic subtraction terms, as shown in the Appendix of \cite{Ferreiro:2022ibf}). Our scheme yields, for a light scalar field in de Sitter space ($m \ll H$), a scale-invariant regularized power spectrum $\Delta_{\phi}^{\rm (reg)} \simeq H^2 /(4\pi^2)$ at super-Hubble scales.

In this work we continue the program initiated in Ref.~\cite{Ferreiro:2022ibf}, by extending the method to the regularization of the scalar field's stress-energy tensor. As we shall see, in this case the subtraction terms will depend instead on two arbitrary mass scales $\Mma$ and $\Mmb$, which minimize the infrared distortions if they take sufficiently large values. As we shall see, their minimum required value to overcome the problem of the infrared distortions is around the physical scale of the studied scenario, e.g.~the Hubble parameter for de Sitter inflation or the inflaton's oscillation frequency for geometric reheating (see Sect.~\ref{sec:applications} for more details). For this reason, we have named our proposed regularization scheme \textit{physical scale adiabatic regularization} (or `PSAR').

The structure of the paper is as follows. In Sect.~\ref{sec:ScalarEqs} we present the different equations describing the dynamics of scalars field in cosmological spacetimes and review the standard adiabatic regularization method. In Sect.~\ref{sec:Regularization} we present our proposed PSAR method and apply it to the regularization of the stress-energy tensor and the two-point function. In Sect.~\ref{sec:RenormCond} we interpret our regularization scheme in terms of renormalization of coupling constants, which we then fix according to a physically motivated renormalization condition. In Sect.~\ref{sec:applications} we illustrate our proposed regularization method in two scenarios of cosmological interest: de Sitter expansion and geometric reheating after inflation. In Sect.~\ref{sec:Summary} we discuss our results and conclude.

\section{Scalar field in a cosmological spacetime} \label{sec:ScalarEqs}

The aim of this work is to reexamine the regularization of expectation values built from free scalar fields propagating in cosmological spacetimes. Let us first introduce the basic equations describing their dynamics. The action of a scalar field $\phi$ can be written as,
\be S_{\phi}[\phi,g_{\mu\nu}] = \frac12 \int d^4x \sqrt{-g} \left(  g^{\mu \nu} \partial_{\mu} \phi \partial_{\nu} \phi -( \xi R+m^2) \phi^2\right) \ , \label{eq:scalar-action} \ee
where $g$ is the determinant of the spacetime metric $g_{\mu \nu}$ and $m^2$ is the mass squared of the field. We have included a non-minimal coupling between the field and the Ricci scalar $R$, whose strength is parametrized by the dimensionless constant $\xi$. The equation of motion is
\be (\Box+\xi R+m^2)\phi=0\ , \hspace{0.5cm}  [\Box \equiv \nabla^{\mu} \nabla_{\mu} ] \ ,  \label{KGphi} \ee
and its stress-energy tensor can be written as ($a,b,c=0,1,2,3$)
\begin{align}
T_{ab} &\equiv 2 |g|^{-1/2} \frac{\delta S_{\phi} }{\delta g^{a b} } \\*
& = \nabla_a\phi\nabla_b\phi-\frac12 g_{ab}\nabla^c\phi\nabla_c\phi+g_{ab}\frac{m^2}{2} \phi^2-\xi\left(R_{ab}-\frac12 R g_{ab}\right)\phi^2+\xi \left(g_{ab}\nabla^c\nabla_c\phi^2-\nabla_a\nabla_b\phi^2\right) , \nonumber
\end{align}
In the particular case of the spatially-flat FLRW metric $ds^2=a^2(\tau)(d \tau^2 - d {\bf x}^2)$ (where $\tau$ denotes the conformal time coordinate), Eq.~\eqref{KGphi} reads as (${}' \equiv d / d \tau$)
\begin{align}
 \phi '' + 2 \frac{a'}{a} \phi' - \nabla^2 \phi + m^2 \phi + \xi R (\tau) \phi= 0 \ , \hspace{0.5cm} R (\tau) = 6 \frac{a''}{a^3} \ . \label{eq:scalar-eom} \end{align}

Let us now quantize our scalar field. We do this by upgrading the field to an operator and performing the following decomposition in terms of field modes,
\be
    \phi(\textbf{x},\tau)=\frac{1}{(2\pi)^3}\int \frac{d^3 \mathbf{k}}{a(\tau)} \left( A_{\bf k} \chi_k(\tau) e^{i {\bf k} \cdot {\bf x}}+A_{\bf k}^{\dagger} \chi_k^*(\tau) e^{-i {\bf k} \cdot {\bf x}}\right) \ , 
\ee
where $A_{\bf k}^{\dagger}$ and $A_{\bf k}$ are creation and destruction operators obeying the commutation relations $[A_{\bf k},A_{\bf k'}^{\dagger}] = \delta_{{\bf k},{\bf k'}}$. The field modes $\chi_k$ are solutions of the equation of motion,
\be
\chi_k''+\left(k^2+m^2a^2+\left(\xi-\frac16\right)a^2 R\right)\chi_k=0 \ ,  \label{eq:fieldmodeeq}
\ee
and must also obey the following normalization condition in order to ensure the standard canonical commutation relations:
\be \chi_k{\chi_k^*}'-\chi_k'\chi_k^*=i \ . \label{eq:normalization} \ee 
A particular solution for $\chi_k$ defines a vacuum state $\ket{0}$ through the condition  $A_{\bf k} \ket{0} = 0$. We can then use the above decomposition to build the two-point function between two spacetime points $x=(\tau, {\bf x})$ and $x'=(\tau', {\bf x}')$ as follows,
\be
\langle \phi(x) \phi(x')\rangle \equiv \langle 0 |\phi(x) \phi(x') |0 \rangle  =\frac{1}{(2\pi)^3}\int d^3k \, \frac{\chi_k(\tau)}{a(\tau)}\frac{\chi^*_k(\tau')}{a(\tau')}e^{i{\bf k}({\bf x}-{\bf x}')}. \label{unrG} 
\ee
We are concerned with the regularization of this quantity at coincident spacetime points $x \rightarrow x'$, as in this limit it contains quadratic and logarithmic ultraviolet divergences. Similarly, we can express the expectation value of the scalar field's stress-energy tensor in the following perfect fluid form:
\begin{align}
    \langle T_{ab}\rangle \equiv \langle 0 |T_{ab}(x)|0 \rangle =-g_{ab}\langle p\rangle+(\langle p\rangle+\langle \rho\rangle )u_a u_b \ , \label{Trhop}
\end{align}
where $u^a=(a^{-1},0,0,0)$ is the four-velocity of a comoving observer, and $\rho$ and $p$ are the energy and pressure densities of the fluid respectively. In general, both expectation values $\langle \rho\rangle$ and $\langle p\rangle$ contain quartic, quadratic, and logarithmic ultraviolet divergences.

Let us now write expressions for $\langle \phi^2 \rangle$, $\langle \rho \rangle$ and $\langle p \rangle$ in terms of integrals over momenta. By defining conformally rescaled field mode amplitudes as $h_k \equiv \chi_k /a$, we obtain the following expressions:
\begin{align}
\langle \phi^2 \rangle &=  \frac{1}{(2\pi)^3}\int d^3k \langle \phi_k^2 \rangle \ , \hspace{0.5cm} \langle \phi_k^2 \rangle \equiv |h_k|^2 ; \\[8pt]
    \langle \rho \rangle& =  \frac{1}{(2\pi)^3}\int d^3k \langle \rho_k\rangle \ ,  \label{rhok}\\ 
   & \hspace{0.3cm} \langle \rho_k\rangle \equiv  \frac{1}{2a^2} \left\{|h'_{k}|^2+\left(k^2+m^2a^2\right)|h_{k}|^2+6\xi\left(\frac{a'^2}{a^2}|h_{k}|^2+\frac{a'}{a}(h_{k}h'^{*}_{k}+h^{*}_{k}h'_{k})\right)\right\} \ ; \nonumber \\[8pt]
\langle p \rangle &=  \frac{1}{(2\pi)^3}\int d^3k \langle p_k\rangle \ , \nonumber \\
 & \hspace{0.3cm} \langle p_k\rangle \equiv  \frac{1}{2a^2} \left\{ |h'_{k}|^2-\left(\frac{k^2}{3}+m^2a^2\right)|h_{k}|^2-2\xi\left(\left(2-12\xi\right)\frac{a''}{a}-\frac{a'^2}{a^2}\right)|h_{k}|^2 \right.\nonumber \\
& \hspace{1.3cm} \left.+2\xi\left(\frac{a'}{a}(h_{k}h'^{*}_{k}+h^{*}_{k}h'_{k})-2|h'_{k}|^2+\left(2k^2+2m^2a^2\right)|h_{k}|^2\right)\right\} \ . \label{pk}
\end{align}
These expressions allow us to define \textit{unregularized power spectra} $\Delta_{c}$ (for $c = \phi^2, \rho, p$) associated to these quantities as follows:
\begin{alignat}{4}
    \langle \phi^2\rangle =& \int d \, \text{log}\,k \, \Delta_{\phi^2} (k) \ ,  \hspace{0.7cm}  &\Delta_{\phi^2} (k) &\equiv\, & \frac{k^3}{2 \pi^2}  \langle \phi_k^2 \rangle \ , \label{tpf}  \\
    \langle \rho \rangle=&  \int d \, {\rm log} \, k \,
     \Delta_{\rho} (k) \ , \hspace{0.7cm} &\Delta_{\rho} (k) &\equiv \, & \frac{k^3}{2 \pi^2}  \langle \rho_k\rangle \ , \label{rhopower} \\
     \langle p \rangle=&  \int d \, {\rm log} \, k \,
     \Delta_p (k) \ , \hspace{0.7cm} &\Delta_p (k) &\equiv \,& \frac{k^3}{2 \pi^2} \langle p_k\rangle \ , \label{ppower}
\end{alignat}
where we have used $d^3 k \equiv 4 \pi k^3 d \log k$ due to isotropy. 

Equations \eqref{tpf}, \eqref{rhopower} and \eqref{ppower} are divergent in the ultraviolet, so a regularization method needs to be performed in order to cancel these divergences and obtain finite quantities. We can achieve this by subtracting an appropriately chosen set of subtraction terms $\mathcal{S}_c$ (for $c = \phi^2, \rho, p$) inside the momentum integrals. We denote the resulting regularized expectation values as $\langleRE c \rangleRE$. We can then write
\begin{alignat}{5}
   \langleRE \phi^2 \rangleRE &\equiv\frac{1}{(2\pi)^3}\int d^3k\left( \langle \phi^2_k\rangle-\mathcal{S}_{\phi^2}\right)   \hspace{0.5cm}
   &\longrightarrow&\hspace{0.62cm}    
   \SprPhi (k) &\,\equiv\,& \frac{k^3}{2 \pi^2} \left( \langle \phi^2_k\rangle-\mathcal{S}_{\phi^2}\right) , \label{eq:reg-tpf}  \\
   \langleRE \rho \rangleRE & \equiv \frac{1}{(2\pi)^3}\int d^3k\left( \langle \rho_k\rangle-\mathcal{S}_{\rho}\right)  \hspace{0.5cm}
   &\longrightarrow&\hspace{0.7cm}  
   \SprRho (k) &\,\equiv\,& \frac{k^3}{2 \pi^2} \left( \langle \rho_k\rangle-\mathcal{S}_{\rho}\right) , \label{eq:reg-ene} \\ 
    \langleRE p \rangleRE & \equiv \frac{1}{(2\pi)^3}\int d^3k\left( \langle p_k\rangle-\mathcal{S}_{p}\right)  \hspace{0.5cm}
    &\longrightarrow&\hspace{0.7cm}    
    \SprPre (k) &\,\equiv\,& \frac{k^3}{2 \pi^2} \left( \langle p_k\rangle-\mathcal{S}_{p}\right) \label{eq:reg-pre} \ ,
\end{alignat}
where for each expectation value we have defined its \textit{regularized power spectrum} $\SprGen$. If the chosen subtraction terms have the appropriate ultraviolet behavior, they should cancel the UV divergences of the unregularized quantity, and hence the resulting regularized spectra should behave as $\SprGen \sim k^{-\alpha}$ with $\alpha > 0$ for large $k$. Note that the regularized stress-energy tensor can be built from the regularized energy and pressure densities as 
\be \langleRE T_{ab}\rangleRE =-g_{ab}\langleRE  p\rangleRE +(\langleRE  p\rangleRE +\langleRE \rho\rangleRE )u_a u_b \ . \ee

\subsection{Adiabatic regularization} \label{sec:adiabatic}

Let us now review the adiabatic regularization method for scalar fields \cite{Parker:1968mv,Parker:1969au,Parker:1971pt}, which allows us to obtain a set of subtraction terms with the appropriate ultraviolet behavior. The method is based on a WKB-like adiabatic expansion of the field modes, which up to $n$th order takes the form
\begin{align}
    \chi_k^{(n)} \equiv a \, h_k^{(n)} \sim \frac{1}{\sqrt{2W_k^{(n)}(\tau)}}e^{-i\int^\tau W_k^{(n)}(\tau') d\tau } \ , \hspace{0.4cm} W_k^{(n)}\equiv\omega_k^{(0)}+\omega_k^{(1)}+...+\omega_k^{(n)} \ .  \label{WKBansatz}
\end{align}
The superscripts in the different $ \omega_k^{(j)}$ terms (with $j = 0,\dots n$) indicate their adiabatic order, defined as the number of time derivatives of the scale factor they contain (e.g.~$a'''$ is of order three, while $a'^2 a''$ is of order four). We fix the zeroth order of the expansion to $\omega_k^{(0)}=\omega \equiv \sqrt{k^2+m^2a^2}$. By substituting \eqref{WKBansatz} into \eqref{eq:fieldmodeeq} and solving the equation iteratively for increasingly higher adiabatic orders, we can derive the expansion up to any finite order $n$, obtaining this way a unique expression of $\chi_k^{(n)}$. The first terms of the expansion are
\begin{align}
\omega_k^{(0)} &= \omega \ , \label{eq:adsub-0} \\
\omega_k^{(1)} &= \omega_k^{(3)} = 0 \ , \\ 
\omega_k^{(2)} &= \frac12 \omega^{-1/2}\frac{d^2}{d\tau^2}\omega^{-1/2}+ \frac12 \omega^{-1}\left(\xi-\frac16\right)a^2R \ , \\
\omega_k^{(4)} &= \frac14 \omega^{(2)} \omega^{-3/2}\frac{d^2}{d\tau^2}\omega^{-1/2}- \frac12 \omega^{-1}(\omega_k^{(2)})^2-\frac14 \omega^{1/2}\frac{d^2}{d\tau^2}\left(\omega^{-3/2}\omega^{(2)}\right) \label{eq:adsub-4} \ .
\end{align}

In order to obtain the subtraction terms that regularize an expectation value, we must first write it as an integral of an expression of field modes over momenta, like \eqref{eq:reg-tpf}-\eqref{eq:reg-pre}. We then replace the modes $\chi_k$ in such an expression by the expansion \eqref{WKBansatz}, and expand it up to the maximum order that guarantees ultraviolet convergence. Due to dimensional reasons, the stress-energy tensor requires an adiabatic expansion up to fourth order, while the two-point function only requires an expansion up to second order. The adiabatic subtraction terms for the two-point function are
\begin{align}
    \mathcal{S}_{\phi^2}= \left( \langle \phi_k^2 \rangle [\chi_k^{(2)}]  \right)^{(0-2)} = \left( \frac{1}{2 a^2 W_k} \right)^{(0-2)} =  \frac{1}{2a^2\omega}-\frac{\left(\xi-\frac16\right) R}{4\omega^3}-\frac{3}{16}\frac{\omega'^2}{a^2\omega^5}+\frac{\omega''}{8a^2\omega^4}.\label{Q1}
\end{align} 
where the superscript `${(0-n)}$' over a given quantity indicates that orders from $0$ to $n$ of its expansion are included. Similarly, the subtraction terms for the energy and pressure densities can be obtained as\footnote{Here we do not write explicit expressions for these subtraction terms, but we can provide a Mathematica notebook containing them upon request.}
\be
    \mathcal{S}_{\rho}\equiv \left(\langle\rho_k\rangle[\chi_k^{(4)}]\right)^{(0-4)} , \hspace{0.6cm}
    \mathcal{S}_{p} \equiv \left(\langle p_k\rangle[\chi_k^{(4)}]\right)^{(0-4)} \ .\label{subm1}
\ee
One can check that the resulting regularized stress-energy tensor is conserved, and that it vanishes in the limit of Minkowski spacetime (which we denote as $\mathcal{M}$) for the Poincaré-invariant vacuum state, i.e.,  $\langleRE T_{\mu \nu} \rangleRE |_{\mathcal{M}} = 0$.

Before moving on, it is convenient to define a \textit{zeroth-order-subtracted} power spectrum for each quantity, in which only the zeroth order adiabatic term has been subtracted from the unregularized expression, i.e.,
\be \Delta_{c}^{\rm (0s)} (k) \equiv \frac{k^3}{2 \pi^2} \left( \langle c_k\rangle-\mathcal{S}_{c}^{(0)} \right) \ , \hspace{0.5cm} c = \rho, p, \phi^2 \ . \label{eq:MinkSubsPW} \ee
Note that in Minkowski spacetime we have $\Delta_{c}^{\rm (0s)}|_{\mathcal{M}} = 0$, i.e.~this construction successfully removes the vacuum contribution. However, Eq.~\eqref{eq:MinkSubsPW} is still ultraviolet divergent in generic FLRW spacetimes: the one of the two-point function still has a residual logarithmic divergence, while the ones of the stress-energy tensor still have both quadratic and logarithmic ones.

\section{Regularization without infrared distortions} \label{sec:Regularization}

In this paper we wish to construct a regularization scheme for the stress-energy tensor
that removes its ultraviolet divergences, while simultaneously fulfilling the following three conditions:
\begin{itemize}
\item[(i)] It must yield $\langleRE  T_{ab} \rangleRE |_{\mathcal{M}}=0$ in the Minkowskian limit. In other words, the regularization scheme must be equivalent to the well-known normal-ordering prescription in Minkowski spacetime.\vspace{-0.1cm}
\item[(ii)] The  regularized stress-energy tensor must be conserved, i.e.~$\nabla^a\langleRE T_{ab} \rangleRE=0$.\vspace{-0.1cm}
\item[(iii)] The regularization procedure must not significantly distort the amplitude of the power spectra \eqref{eq:reg-ene} and \eqref{eq:reg-pre} at the momenta scales amplified by the non-adiabatic expansion of the universe. More specifically, we require that $\SprGen \hspace{-0.05cm} (k)\approx\Delta_{c}^{\rm (0s)} (k)$ (with $c=\rho, p$) for all infrared modes $k \lesssim k_{+}$, where $k_{+}$ is the maximum amplified momentum.
\end{itemize}

The standard adiabatic scheme reviewed in Sect.~\ref{sec:adiabatic} does indeed satisfy conditions (i) and (ii). However, it fails to satisfy condition (iii). The reason is apparent if we inspect e.g.~the subtraction terms for the two-point function \eqref{Q1}: for momenta $k \gtrsim a m$, contributions to the momentum integral from second- and higher-order terms of the expansion behave as $\sim k^2 \cdot \omega^{-q/2} \sim k^{2 -q}$ with $q \gtrsim 3$. Therefore, if the maximum amplified physical momentum is larger than the field mass $m a \lesssim k_+$, the subtractions will introduce spurious infrared distortions. We shall see an example of this in Sect.~\ref{sec:deSitter}, where we consider a field with mass $m \ll H$ (where $H$ is the Hubble parameter) propagating in de Sitter spacetime. In this scenario we have $k_+ \approx H$, so the adiabatic scheme does indeed introduce unwanted distortions.

The root of the problem lies in the zeroth-order term of the expansion, which in the standard adiabatic approach is set to $\omega_k^{(0)} = \omega = (k^2 + a^2 m^2)^{1/2}$. It is thus desirable to construct a regularization scheme that incorporates the advantages of adiabatic regularization but also gets rid of this assumption, so that we are able to reproduce the spectrum at infrared scales. A first step in this direction was taken by some of us in \cite{Ferreiro:2022ibf}, where we developed such a construction for the regularized two-point function at coincident spacetime points. Here we generalize our results to the more complicated regularized stress-energy tensor.

Note that an important result in renormalization theory in curved spacetime is that two different regularization methods for the stress-energy tensor compatible with locality and covariance (or more generally satisfying the Wald axioms \cite{Wald:1995yp}) can differ by a finite amount of geometrical terms
\be  \langleRE T_{ab}\rangleRE - \widetilde{\langleRE T_{ab}\rangleRE } =\alpha g_{ab}+\beta G_{ab}+\gamma \,\, \HOne \ , \label{arbT} \ee
where $\{\alpha,\beta,\gamma\}$ are three dimensionless parameters, $G_{ab}$ is the Einstein tensor, and $\HOne=2R_{;ab}-2g_{ab}\Box R-\frac12 g_{ab}R^2+2RR_{ab}$ is a higher-order geometrical tensor\footnote{Note that if we were working in a general curved background, we would need to include a second higher-order tensor $\HTwo= R_{;ab}-\frac12g_{ab}\Box R-\Box R_{ab}-\frac12 g_{ab}R_{ab}R^{ab}+2R^{cd}R_{cdab}$ in \eqref{arbT}. However, $\HTwo$ is proportional to $\HOne$ in a FLRW spacetime so it is not independent.}.

We can make use of this arbitrariness to construct an alternative subtraction scheme that is equivalent to adiabatic regularization up to the geometrical terms in \eqref{arbT}. Since adiabatic regularization has been shown to be equivalent to methods in general curved spacetimes \cite{delRio:2014bpa,Beltran-Palau:2021xrk}, the new method will also be equivalent to these up to geometric arbitrariness. In the following section we show how to incorporate the allowed arbitrariness in the subtraction terms of $\langle \rho \rangle$, $\langle p \rangle$ and $ \langle \phi^2 \rangle$.

\subsection{Construction of the physical scale adiabatic regularization (PSAR) method}

We now construct an alternative regularization method that fulfills all three conditions of the above list, using as a basis the arbitrariness allowed by Eq.~\eqref{arbT}. Our starting point is the work carried out in Ref.~\cite{Ferreiro:2018oxx}, in which the standard adiabatic expansion was generalized with the introduction of an arbitrary mass scale $\mu$, obtaining this way a generalized `off-shell' type of prescription for adiabatic regularization. We denote this new approach as $\mu$-\textit{adiabatic} regularization. In order to illustrate this method, it is convenient to rewrite the field mode equation \eqref{eq:fieldmodeeq} as follows,
\begin{align}
    \chi_k''+\left(\omega_{\mu}^2-\mu^2a^2+m^2a^2+\left(\xi-\frac16\right)a^2 R\right)\chi_k=0 \ , \hspace{0.5cm} \omega_{\mu} \equiv \sqrt{k^2+\mu^2a^2} \ ,\label{eq:fieldmodeeq2}
\end{align}
where the mass parameter $\mu$ has been introduced. As in the standard adiabatic procedure, we expand the field modes with the WKB-like template
\begin{align}
    \overline{\chi}_k^{(n)} \equiv a \, \overline{h}_k^{(n)} \sim \frac{1}{\sqrt{2\overline{W}_k^{(n)}(\tau)}}e^{-i\int^\tau \overline{W}_k^{(n)}(\tau') d\tau } \ , \hspace{0.4cm} \overline{W}_k^{(n)}\equiv\overline{\omega}_k^{(0)}+\overline{\omega}_k^{(1)}+...+\overline{\omega}_k^{(n)} \ ,  \label{WKBansatz2}
\end{align}
but now the zeroth order of the expansion is fixed to be $\overline{\omega}_k^{(0)}=\omega_{\mu}$ (the overline notation will be used to denote terms constructed from this generalized $\mu$-adiabatic expansion). Higher-order terms of the expansion can be obtained by substituting the ansatz \eqref{WKBansatz2} into \eqref{eq:fieldmodeeq2} and solving order by order, where the rest of the parentheses of \eqref{eq:fieldmodeeq2} must be considered of adiabatic order two (i.e.~in this expansion both $m^2$ and $\mu^2$ are of order two). We obtain
\begin{align}
\overline{\omega}_k^{(0)} &= \omega_{\mu} \ , \\\hspace{0.4cm}
\overline{\omega}_k^{(1)} &= \overline{\omega}_k^{(3)} = 0 \ , \\\hspace{0.4cm}
\overline{\omega}_k^{(2)} &=\frac12 \omega_{\mu}^{-1/2}\frac{d^2}{d\tau^2}\omega_{\mu}^{-1/2}+ \frac12 \omega_{\mu}^{-1}\left[\left(\xi-\frac16\right)a^2R+a^2m^2-a^2\mu^2\right] \ , \\
\overline{\omega}_k^{(4)} &=\frac14 \overline{\omega}_k^{(2)} \omega_{\mu}^{-3/2}\frac{d^2}{d\tau^2}\omega_{\mu}^{-1/2}- \frac12 \omega_{\mu}^{-1}(\overline{\omega}_k^{(2)})^2-\frac14 \omega_{\mu}^{1/2}\frac{d^2}{d\tau^2}\left(\omega_{\mu}^{-3/2}\overline{\omega}_k^{(2)}\right) \ . 
\end{align}
Note that these terms coincide with the standard adiabatic ones \eqref{eq:adsub-0}-\eqref{eq:adsub-4} for the choice $\mu = m$. 

The $\mu$-dependent subtraction terms for the two-point function are obtained, as before, by expanding $\langle \phi_k^2 \rangle$ up to second order,
\be
\overline{ \mathcal{S}_{\phi^2}} \equiv \overline{\langle \phi_k^2 \rangle}^{(0-2)} = \frac{1}{2a^2\omega_{\mu}}+\frac{(\mu^2-m^2)}{4\omega_{\mu}^3}-\frac{(\xi-\frac16)R }{4\omega_{\mu}^3} -\frac{3}{16}\frac{\omega'^{2}_{\mu}}{a^2 \omega_{\mu}^5} +\frac{\omega''_{\mu}}{8a^2\omega_{\mu}^4} \ . \label{eq:two-pf-ext}
\ee
Similarly, one could in principle construct subtraction terms for the stress-energy tensor by simply expanding $\langle \rho_k \rangle^{(0-4)}$ and $\langle p_k \rangle^{(0-4)}$ up to fourth order. However, one can show that the regularized stress-energy tensor obtained this way is not conserved. We can solve this issue by modifying the subtraction terms as follows:
\begin{align}
\mathcal{\overline S}_{\rho}&\equiv\overline{\langle \rho_k \rangle}^{(0-4)}+\frac{\mu^2}{2a^2}\overline{\langle \phi_k^2\rangle}^{(4)} \ ,  \label{eq:Srhomu} \\*
\mathcal{\overline S}_p&\equiv\overline{\langle p_k \rangle}^{(0-4)}-\frac{(1-4\xi)\mu^2}{2a^2}\overline{\langle\phi_k^2\rangle}^{(4)} \ , \label{eq:Spmu} 
\end{align}
where the extra terms in $\overline{\mathcal{S}_{\rho}}$ and $\overline{\mathcal{ S}_p}$ are chosen such that the condition $\nabla^a \overline{\langleRE T_{ab} \rangleRE}=0$ holds. Observe that the subtraction terms $\overline{ \mathcal{S}_{\phi^2}}$, $\mathcal{\overline S}_{\rho}$ and $\mathcal{\overline S}_p$ coincide with the standard adiabatic ones $\mathcal{S}_{\phi^2}$, $\mathcal{ S}_{\rho}$ and $\mathcal{S}_p$ for $\mu = m$.

Note that some of the fourth-order subtraction terms obtained with \eqref{eq:Srhomu} and \eqref{eq:Spmu} will be finite after integration in momenta and can be integrated out. They are hence not strictly necessary in order to remove the ultraviolet divergences of the stress-energy tensor. In that regard, it is convenient to decompose the subtraction terms as $\overline{S}_{c} = \overline{S}_{c}^{\rm (d)} + \overline{S}_{c}^{\rm (f)}$ for $c=\rho, p$, where $\overline{S}_{c}^{\rm (d)} $ contains all the divergent subtraction terms, and $\overline{S}_{c}^{\rm (f)}$ is composed by the finite terms that can be expressed as geometric tensors after integration in momenta. The regularized stress-energy tensor can then be written as,
\begin{align}
\overline{ \langleRE T_{ab} \rangleRE }& =\int d\log k\left(-(\Delta_{p}-\mathcal{\overline S}_p)g_{ab}+(\Delta_{p}-\mathcal{\overline S}_p+\Delta_{\rho}-\mathcal{\overline S}_{\rho})u_{a}u_b\right)\label{Tabreg} \\&=\int d\log k\left(-(\Delta_{p}-\mathcal{\overline{S}}_p^{\rm (d)})g_{ab}+(\Delta_{p}-\mathcal{\overline{S}}_p^{\rm (d)}+\Delta_{\rho}-\mathcal{\overline{S}}_{\rho}^{\rm (d)})u_{a}u_b\right)+T^{\rm  (f) }_{ab}+\frac{\left(\xi-\frac16\right)}{288\pi^2} \HOne \ , \nonumber \hspace*{-0.3cm}
\end{align}
where in the second line we have defined the following geometric tensor,
\begin{align}
    T^{\rm (f)}_{ab} \equiv \frac{1}{64\pi^2}\left(\frac{1}{45}R_{ac}R^{c}_b-\frac{1}{45}RR_{ab}-\frac{1}{135}\nabla_a\nabla_bR-\frac{1}{90}R_{cd}R^{cd}g_{ab}+\frac{1}{135}R^2g_{ab}+\frac{1}{135}\Box R g_{ab}\right) . \,\,\,\, \hspace{-0.5cm}
\end{align}

Let us explicitly write the divergent part of the subtraction terms. For the energy density we obtain the following sum of zeroth-, second-, and fourth-order terms,
\begin{align}
\mathcal{\overline{S}}^{({\rm d})}_{\rho} \equiv & \,\,\, \mathcal{\overline{S}}^{({\rm d},0)}_{\rho} +\mathcal{\overline{S}}^{({\rm d},2)}_{\rho} + \mathcal{\overline{S}}^{({\rm d},4)}_{\rho} \ , \\
&\mathcal{\overline{S}}^{({\rm d},0)}_{\rho}=-\frac{\mu ^2}{4 \omega_{\mu}  a^2}+\frac{\omega_{\mu} }{2 a^4}+\frac{m^2}{4 \omega_{\mu}  a^2}-\frac{\mu ^4}{16 \omega_{\mu} ^3}+\frac{\mu ^2
   m^2}{8 \omega_{\mu} ^3}-\frac{m^4}{16 \omega_{\mu} ^3} \ , \label{eq:Crho0} \\
&\mathcal{\overline{S}}^{({\rm d},2)}_{\rho}=\left(\xi-\frac16\right) 
   \left(-\frac{9 \mu ^4 a'^2}{4 \omega_{\mu} ^5 a^2}-\frac{3 \mu ^2 a'^2}{4 \omega_{\mu} ^3 a^4}-\frac{3 a'^2}{2 \omega_{\mu}  a^6}+\frac{9 \mu
   ^2 m^2 a'^2}{4 \omega_{\mu} ^5 a^2}-\frac{3 m^2 a'^2}{4 \omega_{\mu} ^3
   a^4}\right) \ , \label{eq:Crho2}  \\
&\mathcal{\overline{S}}^{({\rm d},4)}_{\rho}=\left(\xi-\frac16\right)^2 \left(\frac{27 \mu ^2 a'^2 a''}{2 \omega_{\mu} ^5 a^5}+\frac{9
   a'^2 a''}{\omega_{\mu} ^3 a^7}-\frac{9 a^{(3)} a'}{2
   \omega_{\mu} ^3 a^6}+\frac{9 a''^2}{4 \omega_{\mu} ^3 a^6}\right) \ , \label{eq:Crho4}
\end{align}
while for the pressure density we obtain
\begin{align}
\mathcal{\overline{S}}^{({\rm d})}_{p} \equiv & \,\,\, \mathcal{\overline{S}}^{({\rm d},0)}_{p} +\mathcal{\overline{S}}^{({\rm d},2)}_{p} + \mathcal{\overline{S}}^{({\rm d},4)}_{p} \ , \\
   &\mathcal{\overline{S}}^{({\rm d},0)}_p= -\frac{\mu ^6 a^2}{16 \omega_{\mu} ^5}-\frac{\mu
   ^2}{12 \omega_{\mu}  a^2}+\frac{\omega_{\mu} }{6 a^4}+\frac{\mu ^4 m^2 a^2}{8 \omega_{\mu} ^5}-\frac{\mu ^2 m^4 a^2}{16 \omega_{\mu} ^5}-\frac{m^2}{12 \omega_{\mu} 
   a^2}-\frac{\mu ^4}{48 \omega_{\mu} ^3}-\frac{\mu ^2 m^2}{24 \omega_{\mu}
   ^3}+\frac{m^4}{16 \omega_{\mu} ^3} , \\
   &\mathcal{\overline{S}}^{({\rm d},2)}_p= \left(\xi-\frac16\right)  \left(-\frac{15
   \mu ^6 a'^2}{4 \omega_{\mu} ^7}-\frac{3 \mu ^2 a'^2}{4 \omega_{\mu} ^3 a^4}+\frac{3 \mu ^4 a''}{2 \omega_{\mu} ^5 a}+\frac{\mu ^2 a''}{2
   \omega_{\mu} ^3 a^3}-\frac{3 a'^2}{2 \omega_{\mu}  a^6}+\frac{a''}{\omega_{\mu}  a^5}+\frac{15 \mu ^4 m^2 a'^2}{4 \omega_{\mu} ^7}\right.  \nonumber \\
   & \hspace{1.67cm} \left. -\frac{3 \mu ^2
   m^2 a'^2}{2 \omega_{\mu} ^5 a^2} -\frac{3 \mu ^2 m^2 a''}{2 \omega_{\mu} ^5
   a}  -\frac{m^2 a'^2}{4 \omega_{\mu} ^3 a^4}+\frac{m^2 a''}{2
   \omega_{\mu} ^3 a^3}\right) \ ,   \\
   & \mathcal{\overline{S}}^{({\rm d},4)}_p= \left(\xi-\frac16\right)^2 \left(\frac{45 \mu ^4 a'^2 a''}{2 \omega_{\mu} ^7 a^3}+\frac{18
   \mu ^2 a'^2 a''}{\omega_{\mu} ^5 a^5}-\frac{9 \mu ^2 a^{(3)}
   a'}{\omega_{\mu} ^5 a^4}-\frac{27 \mu ^2 a''^2}{4 \omega_{\mu} ^5 a^4}+\frac{12 a'^2 a''}{\omega_{\mu} ^3 a^7}-\frac{15 a^{(3)}
   a'}{2 \omega_{\mu} ^3 a^6}\right. \nonumber \\*
& \hspace{1.67cm} \left.-\frac{15 a''^2}{4 \omega_{\mu} ^3 a^6} 
 +\frac{3 a^{(4)}}{2 \omega_{\mu} ^3 a^5}\right)   \ .
\end{align}

One can show that the difference between the stress-energy tensors regularized with the adiabatic and $\mu$-adiabatic approaches can be written as in Eq.~(\ref{arbT}), where $\alpha$, $\beta$ and $\gamma$ depend on $\mu$. Therefore, both methods satisfy the Wald axioms. More generally, Eq.~(\ref{arbT}) allows us to introduce different arbitrary mass scales at zeroth, second, and fourth adiabatic orders, which we denote as $M_0$, $M_2$ and $M_4$ respectively. We can then change the subtraction terms as follows,
\begin{align}
&\{\mathcal{\overline{S}}^{(d,0)}_{\rho}(\mu),\mathcal{\overline{S}}^{(d,0)}_{p}(\mu)\} \,\,\,\,\,\, \to \,\,\,\,\,\,  \{\mathcal{\overline{S}}^{(d,0)}_{\rho}(M_0),\mathcal{\overline{S}}^{(d,0)}_{p}(M_0)\} \ , \\*
&
\{\mathcal{\overline{S}}^{(d,2)}_{\rho} (\mu),\mathcal{\overline{S}}^{(d,2)}_{p}(\mu)\} \,\,\,\,\,\, \to  \,\,\,\,\,\,  \{\mathcal{\overline{S}}^{(d,2)}_{\rho} (M_2),\mathcal{\overline{S}}^{(d,2)}_{p}(M_2)\} \ ,\\* 
& \{\mathcal{\overline{S}}^{(d,4)}_{\rho}(\mu),\mathcal{\overline{S}}^{(d,4)}_{p}(\mu)\} \,\,\,\,\,\,  \to  \,\,\,\,\,\,  \{\mathcal{\overline{S}}^{(d,4)}_{\rho}(M_4),\mathcal{\overline{S}}^{(d,4)}_{p}(M_4)\} \ .
\end{align}

Let us now analyze if these subtraction terms fulfill the three conditions listed above. Regarding condition (ii), the resulting regularized stress-energy tensor is conserved by construction. Regarding condition (i), the correct Minkowskian limit is satisfied only if we fix $M_0 = m$ at zeroth order, so we impose that condition from now on. Finally, regarding condition (iii), these terms guarantee that the distortions introduced at infrared scales are minimized for large enough values of $M_2$ and $M_4$ (we will illustrate this in two specific examples in Sect.~\ref{sec:applications}). Therefore, we define our definitive subtraction terms for the regularized stress-energy tensor as follows:
\begin{align}
    \mathcal{\widetilde{S}}^{({\rm d})}_{\rho} =&\,\frac{\omega}{2 a^4}+\left(\xi-\frac16\right) \left(\frac{9 \Mma ^2 m^2
        a'^2}{4 a^2 \omegaa^{\,5}}-\frac{3 m^2 a'^2}{4 a^4 \omegaa^3} -\frac{9 \Mma ^4 a'^2}{4 a^2 \omegaa^5}-\frac{3 \Mma ^2 a'^2}{4 a^4
   \omegaa ^3}-\frac{3 a'^2}{2 a^6 \omegaa}\right) \label{eq:subs-rho} \\
   & + \left(\xi-\frac16\right)^2 \left(\frac{27 \Mmb^2 a'^2 a''}{2 a^5 \omegab^5}+\frac{9 a'^2 a''}{a^7 \omegab^3}-\frac{9 a^{(3)} a'}{2 a^6
  \omegab^3}+\frac{9 a''^2}{4 a^6 \omegab^3}\right) \ ,  \nonumber\\
 \mathcal{\widetilde{S}}^{({\rm d})}_{p}  =&\,-\frac{m^2}{6 a^2 \omega}+\frac{\omega}{6 a^4}+\left(\xi-\frac16\right) \left(\frac{3 \Mma^4 a''}{2 a \omegaa^5}+\frac{\Mma^2 a''}{2 a^3 \omegaa^3}+\frac{a''}{a^5 \omegaa}-\frac{15 \Mma^6 a'^2}{4 \omegaa^7}-\frac{3 \Mma^2 a'^2}{4 a^4 \omegaa^3}-\frac{3 a'^2}{2 a^6 \omegaa}\right)\nonumber\\&+\left(\xi-\frac16\right)\left(-\frac{3 m^2 \Mma^2 a''}{2 a \omegaa^5}+\frac{m^2 a''}{2 a^3 \omegaa^3}+\frac{15 m^2 \Mma^4 a'^2}{4 \omegaa^7}-\frac{3 m^2 \Mma^2 a'^2}{2 a^2 \omegaa^5}-\frac{m^2 a'^2}{4 a^4 \omegaa^3}\right)\nonumber\\&+\left(\xi-\frac16\right)^2
   \left( \frac{3 a^{(4)}}{2 a^5 \omegab^3}-\frac{27 \Mmb^2 a''^2}{4 a^4 \omegab^5}-\frac{15 a''^2}{4 a^6 \omegab^3}-\frac{9 \Mmb^2 a^{(3)} a'}{a^4 \omegab^5}-\frac{15 a^{(3)} a'}{2 a^6 \omegab^3}+\frac{45 \Mmb^4 a'^2 a''}{2 a^3 \omegab^7} \right. \label{eq:subs-pres} \\
   &\left. +\frac{18 \Mmb^2 a'^2 a''}{a^5 \omegab^5} +\frac{12 a'^2 a''}{a^7 \omegab^3}\right) \ , \nonumber 
\end{align}
where we have defined $\omega \equiv \sqrt{k^2+m^2a^2}$, $\omegaa \equiv \sqrt{k^2+\Mma^2a^2}$ and $ \omegab \equiv \sqrt{k^2+\Mmb^2a^2}$.

We name this new regularization method \textit{physical scale adiabatic regularization} (or `PSAR'), since the two extra parameters are conceived to be of order of the physical energy scale of the system we are trying to probe, as we shall see in Sect.~\ref{sec:RenormCond}. This method does not only provide a consistent regularization mechanism, but it is also physical in the sense that it does not act on the infrared amplification of the quantum fluctuations, but only on the ultraviolet part. Note that the difference between the stress-energy tensors regularized with the adiabatic and PSAR schemes can be written as in Eq.~\eqref{arbT}, and hence the PSAR scheme is also compatible with the Wald axioms.

A similar exercise can be carried out for the two-point function at coincident spacetime coordinates. The subtraction terms provided by the $\mu$-regularization prescription have been given in Eq.~\eqref{eq:two-pf-ext}. Analogously to the stress-energy tensor, we set $\mu = m$ in the zeroth-order terms in order to get $\SprPhi|_{\mathcal{M}} = 0$ in the Minkowskian limit, as well as remove the last two subtraction terms because they only provide a finite contribution (see Ref.~\cite{Ferreiro:2022ibf} for details). We then have
\be
\widetilde{\mathcal{S}_{\phi^2}} \equiv \frac{1}{2a^2\sqrt{k^2 + m^2 a^2}}-\frac{(\xi-\frac16)R }{4 (k^2 + M^2 a^2)^{3/2}}\ , \label{eq:two-pf-sub}
\ee
where $M$ is an arbitrary mass scale. Equations \eqref{eq:subs-rho}-\eqref{eq:two-pf-sub} are the most important results of this work.

\section{Renormalization conditions and coupling constants} \label{sec:RenormCond}

We have constructed a regularized stress-energy tensor $ \widetilde{\langleRE T_{ab} \rangleRE_{\bf{M}}}$ that depends on the vector of arbitrary parameters ${\bf{M}} = (\Mma,\Mmb)$ through the subtraction terms \eqref{eq:subs-rho}-\eqref{eq:subs-pres}. However, these parameters are in principle completely arbitrary and hence not physical, so we need to impose some renormalization conditions on $\widetilde{ \langleRE T_{ab}\rangleRE}_{\bf{M}}$ in order to fix their observational value consistently. The quantity $\widetilde{\langleRE T_{ab} \rangleRE}_{\bf{M}}$  couples to the gravitational field via the semiclassical Einstein equations,
\begin{align}
G_{ab}+\Lambda g_{ab}+\alpha \, \HOne=-8\pi G\left( \widetilde{\langleRE T_{ab}\rangleRE}_{\bf{M}}+T_{ab}^{\rm clas}\right) ,
\end{align}
where we have included a term proportional to the second-order curvature tensor $\HOne$ because it is necessary for reabsorbing the divergences of $\widetilde{ \langleRE T_{ab} \rangleRE_{\bf{M}} }$. We have also added an unspecified classical stress-energy tensor $T_{ab}^{\rm clas}$ in the source term for completeness. We wish to impose renormalization conditions so that the coupling constants $\{G,\Lambda,\alpha\}$ in the Einstein equations can be exchanged by their physical values $\{G_{\rm o},\Lambda_{\rm o},\alpha_{\rm o}\}$ observed today.

Let us consider a massive scalar field with mass $m$ such that $\mathcal{R}\ll m^2$ at late times in the evolution of the Universe, where $\mathcal{R}$ includes all possible combinations of curvature tensors (e.g.~$R$, $R_{ab}R^{ab}$, etc). We can then use the adiabatic expansion \eqref{WKBansatz} to approximate the modes $\chi_k$ at these late times. If we fix $\Mma=\Mmb=m$ for now, the subtraction terms in the PSAR scheme coincide with the standard adiabatic ones. It is hence not difficult to show that
\begin{align}
\widetilde{\langleRE T_{ab}\rangleRE}_{\bf{m}} = \langleRE T_{ab} \rangleRE \approx m^{-2}\mathrm{T_{ab}}+\mathcal{O}(m^{-4}) \ , 
\end{align}
where ${\bf m}=(m,m)$ and $\mathrm{T_{ab}}$ is a tensor of adiabatic order six and therefore of dimension six. Therefore, in the late-time regime  $\mathcal{R}\ll m^2$ we can neglect the contribution of the quantum vacuum to the stress-energy tensor, and the semiclassical Einstein's equations reduce to the classical ones used to measure the coupling constants $\{G_{\rm o},\Lambda_{\rm o},\alpha_{\rm o} \}$,
\begin{align}
G_{ab}+\Lambda_{\rm o} g_{ab}+\alpha_{\rm o}\HOne=-8\pi G_{\rm o} T_{ab}^{\rm clas} \ . 
\end{align}
In general we have
\begin{align}
G_{ab}+\Lambda_{\rm o} g_{ab}+\alpha_{\rm o}\HOne=-8\pi G_{\rm o}\left(\widetilde{\langleRE T_{ab} \rangleRE}_{\bf m}+T_{ab}^{\rm clas}\right) \ , \label{eq:Ensteq1}
\end{align}
which is valid also when $\mathcal{R}\gtrsim m^2$ (like in many early-universe scenarios).

We are now interested in incorporating the quantum regularized stress-energy tensor at arbitrary scales $\Mma$ and $\Mmb$. We then need to change the coupling constants adequately,
\be
G_{ab}+\Lambda({\bf{M}}) g_{ab}+\alpha({\bf{M}}) \HOne=-8\pi G({\bf{M}})\left(\widetilde{ \langleRE T_{ab} \rangleRE}_{\bf{M}}+T_{ab}^{\rm clas}\right) \ . \label{eq:Ensteq2}
\ee
The difference between the stress-energy tensors regularized at scales ${\bf M}$ and ${\bf m}$ can be computed to be
\begin{equation}
\widetilde{ \langleRE T_{ab} \rangleRE}_{\bf{M}}- \widetilde{\langleRE T_{ab} \rangleRE}_{\bf m}=\frac{\xi-\frac16}{16\pi^2}\left(m^2-\Mma^2+m^2\log{\left(\frac{\Mma^2}{m^2}\right)}\right)G_{ab}+\frac{(\xi-\frac16)^2}{8\pi^2}\log{\left(\frac{\Mmb^2}{m^2}\right)} \,\HOne \ . \hspace*{1cm} \hspace*{-1cm} \label{eq:STensDiff}
\end{equation}
By subtracting Eq.~\eqref{eq:Ensteq1} from \eqref{eq:Ensteq2} and substituting \eqref{eq:STensDiff} into the resulting expression, we can obtain the following relations between the running couplings constants (which depend on ${\bf M}$) and their observed values today:
\begin{align}
&G(\Mmb)=\frac{G_{\rm o}}{1+\frac{G_{\rm o}(\xi-\frac16)}{2\pi} \left(\Mma^2-m^2-m^2\log{\left(\frac{\Mma^2}{m^2}\right)}\right)} \ , \\
&\Lambda(\Mma)=\frac{\Lambda_{\rm o} }{G_{\rm o}} G(\Mma) \ , \label{eq:LambdaMma} \\
&\alpha(\Mma,\Mmb)=\frac{G(\Mma)}{G_{\rm o}}\left(\alpha_{\rm o}-\frac{(\xi-1/6)^2}{\pi}\log{\left(\frac{\Mmb^2}{m^2}\right)}\right) \ .
\end{align}
The running of the coupling constants must be evaluated on a case-by-case basis. Taking into account that $G_{\rm o} = m_{\rm p}^2$, one can see that the change in the gravitational coupling constant is negligible if $(\xi-\frac16) (\Mma/m_p)^2 \ll 1$, and hence in the cosmological constant $\Lambda$ via Eq.~\eqref{eq:LambdaMma}. For example, in Sect.~\ref{sec:deSitter} we will consider the case of a scalar field with $m \ll H$ and minimal coupling $\xi=0$ in de Sitter space, where $H$ is the Hubble parameter. As we shall see, we need to fix $\Mma \sim H (\ll m_p)$ in order to minimize the infrared distortions, so this condition holds and the change in the coupling constants is negligible. 

The running of the couplings
with scaling $\mu \sim H$ has been widely studied in several works \cite{Moreno-Pulido:2020anb,Moreno-Pulido:2022phq,Moreno-Pulido:2022upl} (including also fermions \cite{Moreno-Pulido:2023ryo}) in the context of the running vacuum model, see \cite{SolaPeracaula:2022hpd} for a review. The possible cosmological consequences have also been studied in \cite{SolaPeracaula:2021gxi,SolaPeracaula:2023swx}.

We can show that the trace of the vacuum expectation value of the stress-energy tensor is given by
\begin{align}
\widetilde{\langleRE T^a_ a \rangleRE}_{\bf M}=&(3(\xi-\frac16) \Box +m^2)\left(\langleRE \phi^2 \rangleRE_{\bf M}-\frac{R}{288\pi^2}\right)+ {\rm Tr}_{a}^a
-\frac{(m^2-\Mma^2)(\xi-1/6)}{16\pi^2}R\nonumber \\
&+\frac{m^2(\xi-1/6)}{8\pi^2}R \log\left(\frac{M}{\Mma}\right)+\frac{3(\xi-1/6)^2}{8\pi^2}\Box R \log\left(\frac{M}{\Mmb}\right) \ , 
\end{align}
where we have defined
\begin{align}
{\rm Tr}_{a}^a \equiv \frac{1}{64\pi^2}\left(-\frac{1}{45}R_{ab}R^{ab}+\frac{1}{135}R^2+\frac{1}{45}\Box R\right)-\frac{(\xi-1/6)^2}{32\pi^2}R^2 \ . \label{traceaa}
\end{align}
One can observe that the relation between the trace of the stress-energy tensor and the two-point function is still maintained at the quantum level, up to renormalization freedom and the contribution of \eqref{traceaa}, which is nothing but the well-known trace anomaly in the conformal coupling limit $\xi=1/6$. 

Finally, let us address an important consequence 
 of result \eqref{eq:STensDiff}. Let us recall that standard adiabatic regularization has been shown to be equivalent to Hadamard/DeWitt Schwinger regularization in four dimensions, see \cite{delRio:2014bpa,Pla:2022spt}. Since the difference between the stress-energy tensors regularized with the PSAR method and thestandard adiabatic one is a sum of covariant tensors, we can affirm that the PSAR method also yields a covariant stress-energy tensor.

\section{Examples} \label{sec:applications}

In this section we apply our proposed PSAR method to two cosmological scenarios of interest: a light scalar field propagating in de Sitter spacetime (see Sect.~\ref{sec:deSitter}), and a scalar field non-minimally coupled to curvature getting excited through a process of geometric reheating after inflation (see Sect.~\ref{sec:geometricreheating}). In these examples we define the regularized power spectrum for the energy and pressure densities as $\SprGen\, \equiv k^3 (\langle c_k \rangle - \mathcal{S}^{\rm (d) }_c ) / (2 \pi^2) $ (where $c = \rho, p$), such that the regularized power spectrum takes the following form [see Eq.~\eqref{Tabreg}]:
\begin{align}
\widetilde{ \langleRE T_{ab} \rangleRE }& =\int d\log k\left(-\SprPre g_{ab}+(\SprPre+ \SprRho)u_{a}u_b\right)+T^{\rm  (f) }_{ab}+\frac{\left(\xi-\frac16\right)}{288\pi^2} \, \HOne \ . \hspace*{-0.3cm}
\end{align}
Note that we do not include the finite contributions from the geometric tensors $T^{\rm  (f) }_{ab}$ and $\HOne$ in the power spectrum definitions. The reason is that we are only interested in the generation of quantum fluctuations/particle production due to the quantum state, while these geometric contributions do not depend on it. These are usually understood as vacuum polarization. The possible consequences of these terms in the inflationary phase have been already investigated in e.g.~\cite{Starobinsky:1980te,Fabris:2000mq}.

\subsection{De Sitter expansion} \label{sec:deSitter}

Let us consider a scalar field propagating on a perfect de Sitter spacetime with constant Hubble parameter $H$. The scale factor evolves as $a(\tau) = - (H \tau)^{-1}$ with  $- \infty < \tau < 0$, and the field mode equation (\ref{eq:fieldmodeeq}) can be written as
\be \chi_k''  + \left( k^2 + \frac{m_H^2 + 12 \xi - 2 }{\tau^2} \right) \chi_k = 0 \ , \hspace{0.5cm} m_H \equiv \frac{m}{H} \label{equbd}\ . \ee
This equation has two independent solutions. Here we consider the Bunch-Davies solution \cite{Bunch:1978yq},
\be \chi_k = \frac{-i\sqrt{\pi \tau}}{2} e^{-\frac{\pi}{2} \mathfrak{Im}(\nu)} H_{\nu}^{(1)} (-k \tau) \ , \hspace{0.5cm} \nu \equiv \sqrt{\frac{9}{4} - m_H^2 - 12 \xi}  \ , \label{eq:fieldsol} \ee
which defines the unique maximally symmetric vacuum state which respect to the underlying symmetries of de Sitter spacetime, satisfies the Hadamard condition \cite{Hollands:2014eia}, recovers the positive-frequency Minkowskian solution for large momenta, and obeys the normalization condition (\ref{eq:normalization})\footnote{Note that the Bunch-Davies vacuum in the massless field case generates a well-known infrared divergence in the two-point function, see e.g.~\cite{Ford:1977in,Allen:1985ux}, so the solution is usually modified to overcome this divergence.}.

Let us now use our PSAR method to compute the regularized power spectra of the energy and pressure densities of the quantized scalar field, defined by the subtraction terms \eqref{eq:subs-rho} and \eqref{eq:subs-pres} respectively. We will set $\xi=0$ for simplicity.  We obtain the following expressions,
\begin{align}
    \SprRho =& \frac{H^4 x^3}{64 \pi^2 } \left(
    \pi e^{- \pi   \mathfrak{Im}(\nu) } \left( 4 (m_H^2 + x^2) |H_{\nu} (x)|^2  + | 3 H_{\nu} (x) +x (H_{\nu - 1} (x) - H_{\nu + 1} (x) )|^2\right) \right. \label{eq:SprRho} \\*
    &  \left. -16 (m_H^2+x^2)^{\frac{1}{2}}
    - \frac{4 \left(6 \MmaH^4+\MmaH^2 \left(5 x^2-2 m_H^2\right)+x^2 \left(m_H^2+2 x^2\right)\right)}{(\MmaH^2+x^2)^\frac{5}{2}}
    - \frac{ 24 \MmbH^2}{ (\MmbH^2+x^2)^{\frac{5}{2}}} \right) \ ,  \nonumber \\ 
    \SprPre =& \frac{H^4 x^3}{192 \pi^2} \left( - \pi e^{- \pi   \mathfrak{Im}(\nu)} \left( 4 (3 m_H^2 + x^2) |H_{\nu}(x)|^2 - 
    3 |3 H_{\nu}(x) + x (H_{\nu - 1}(x) - H_{\nu + 1}(x))|^2\right) \right. \label{eq:SprPre} \\
    & \hspace{-0.8cm} \left.- \frac{16 x^2}{ (m_{H}^2+x^2)^{\frac{1}{2}}} + \frac{4 (20 \MmaH^4 x^2+7 \MmaH^2 x^4+2 x^6)}{ (\MmaH^2+x^2)^{\frac{7}{2}}} + \frac{12 x^2  (\MmbH^2 (10 - 4 m_H^2) + m_H^2 x^2 )}{(\MmbH^2+x^2)^{\frac{7}{2}}} \right) \ ,  \nonumber
\end{align}
where we have defined the dimensionless parameters
\be x \equiv \frac{k}{aH} \ , \hspace{0.5cm} m_H \equiv \frac{m}{H} \ , \hspace{0.5cm} \MmaH \equiv \frac{\Mma}{H} \ , \hspace{0.5cm} \MmbH \equiv \frac{\Mmb}{H} \ . \ee 
Note that the first line in both Eqs.~\eqref{eq:SprRho} and \eqref{eq:SprPre} corresponds to the unregularized spectra, while the second line contains the zeroth-, second-, and fourth-order subtraction terms. These expressions get simplified in the light field limit $m \ll H$,
\begin{align}
    \SprRho & \simeq + \frac{H^4 x^2}{8 \pi ^2} \left( 1 -\frac{6 \MmaH^4 x+5 \MmaH^2 x^3+2 x^5}{2 (\MmaH^2+x^2)^{\frac{5}{2}}}-\frac{3 \MmbH^2 x}{ (\MmbH^2+x^2)^{\frac{5}{2}}} \right) \hspace{1.53cm} [m \ll H] \ ,  \\ 
    \SprPre & \simeq -\frac{H^4 x^2}{24 \pi ^2} \left( 1 - \frac{10 \MmaH^4 x^3 +14 \MmaH^2 x^5+2x^7}{ (\MmaH^2+x^2)^{\frac{7}{2}}} - \frac{15 \MmbH^2 x^3}{(\MmbH^2+x^2)^{\frac{7}{2}}} \right)  \hspace{1cm} [m \ll H] \ ,  \nonumber
\end{align}

\begin{figure*}
    \centering
    \includegraphics[width=0.8\textwidth]{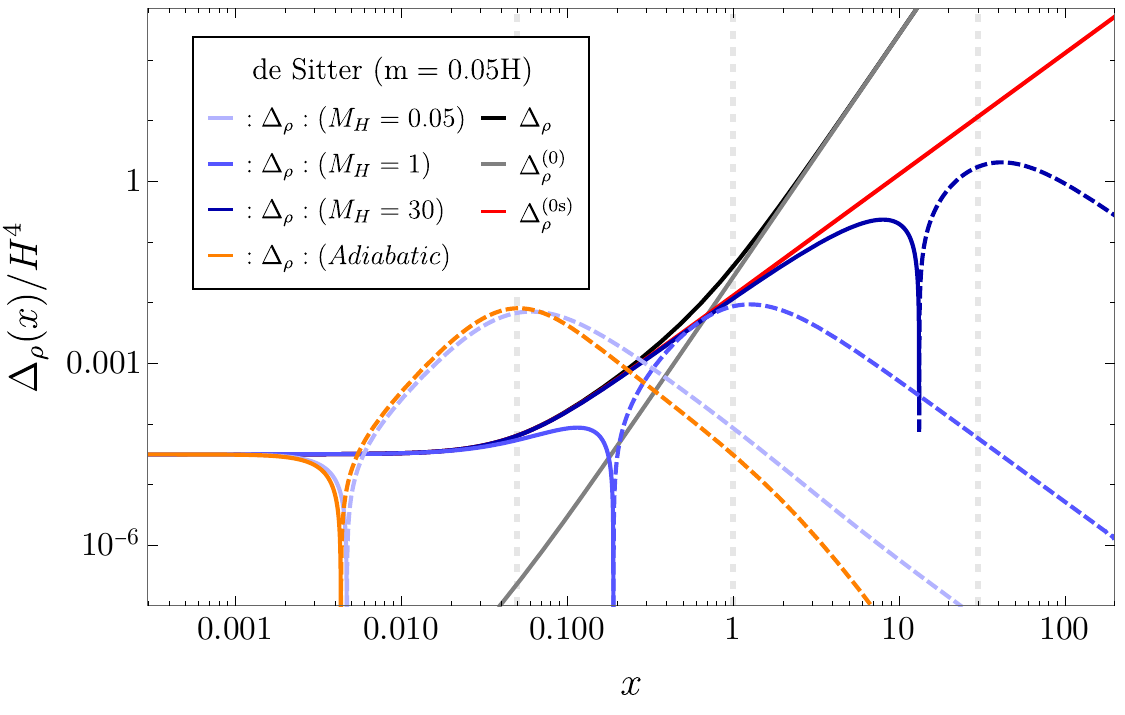} \\ \vspace{0.75cm}
    \includegraphics[width=0.8\textwidth]{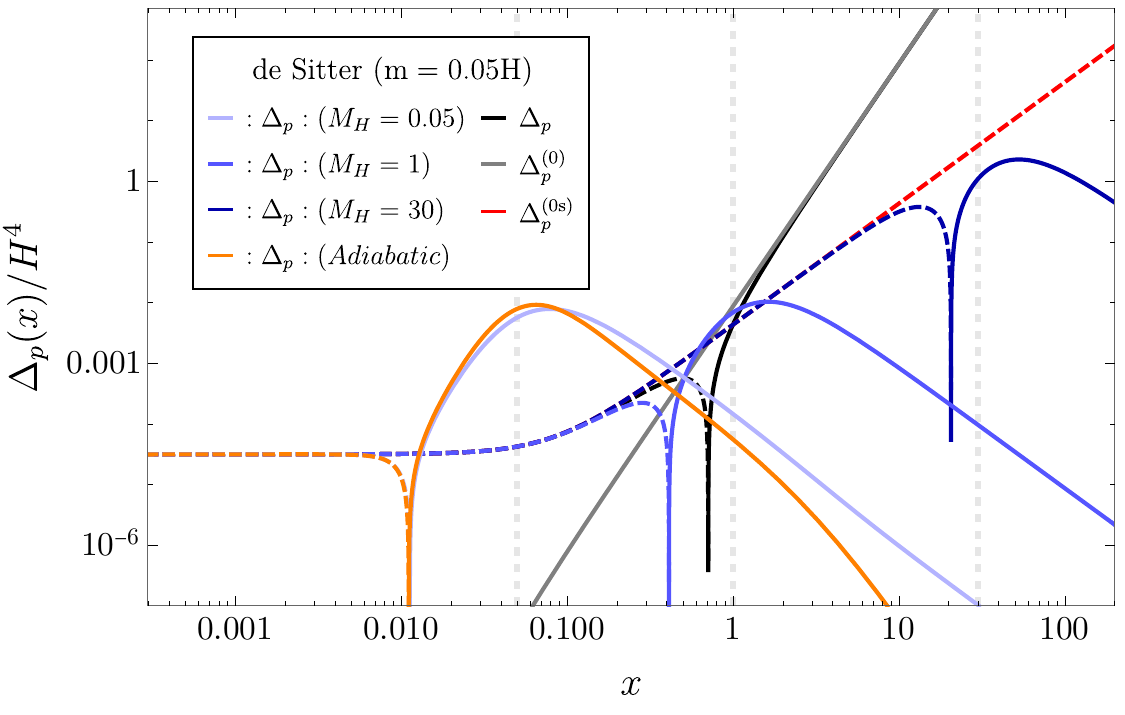}
    \caption{[De Sitter, Sect.~\ref{sec:deSitter}] Power spectra of the energy density (top panel) and pressure density (bottom panel) for a scalar field of mass $m=0.05 H$ in de Sitter space with Hubble parameter $H$. In each panel we show, from top to bottom, the following quantities: the unregularized spectra (black line), the zeroth-order subtraction term (gray), the difference between the two quantities (i.e. the zeroth-order-subtracted spectrum, in red), the spectra regularized with our proposed PSAR scheme for different choices of $\MmaH = \MmbH = M_H$ (blue lines), and the spectra regularized with the adiabatic scheme (orange).}  \label{fig:PowSpecDs}
\end{figure*}

In Fig.~\ref{fig:PowSpecDs} we show the power spectra of the energy and pressure densities (top and bottom panels respectively) for a scalar field with mass $m=0.05H$. The black lines in each panel depict the unregularized spectra, while in gray we depict the zeroth-order subtraction terms $\Delta_c^{(0)} \equiv k^3 \mathcal{S}_{c}^{(0)} / (2 \pi^2)$ (for $c=\rho, p$). The difference between both quantities corresponds to the \textit{zeroth-order-subtracted} spectrum $\Delta_c^{{\rm (0s)}}$ defined in Eq.~\eqref{eq:MinkSubsPW}, and is depicted in red. Note that the quartic ultraviolet divergence of the unregularized spectra gets canceled by the zeroth-order subtraction terms, but $\Delta_c^{{\rm (0s)}}$  still has quadratic and logarithmic divergences.  We can observe that the exponential expansion mainly amplifies super-Hubble modes $x \equiv k/(aH) \lesssim 1$, for which we have $\Delta_{c} / \Delta_c^{(0)} \gg 1$. As explained in Sect.~\ref{sec:Regularization}, we require our regularized spectrum to obey $\SprGen \simeq \Delta_c^{{\rm (0s)}} \equiv \Delta_{c} - \Delta_c^{(0)}$ for those amplified modes, which is guaranteed by our proposed PSAR scheme for large enough values of $\MmaH$ and $\MmbH$.

In order to illustrate this, in Fig.~\ref{fig:PowSpecDs} we plot the PSAR-regularized spectrum $\SprGen$ for the symmetric choice $\MmaH = \MmbH \equiv M_H$ and different values $M_H =0.05, 1, 30$ (blue lines). For comparative purposes, we also plot the power spectrum regularized with the standard adiabatic method (orange line). Note that as expected, the four regularized power spectra are convergent in the ultraviolet, as they behave as $\SprGen \sim k^{-2}$ for large values of $k$. We can clearly see that the PSAR scheme with $M_H = 0.05 (= m_H)$ gives very similar results to the standard adiabatic one. Both regularizations distort the power spectra at the infrared scales $0.005 \lesssim x \lesssim 1$, and in fact, $\SprGen$ has a different sign than $\Delta_c^{{\rm (0s)}}$ at these scales. However, in the PSAR scheme we have an arbitrary parameter $M_H$ that we can arbitrarily increase in order to tame the infrared distortions. The choice $M_H = 1$ only distorts the spectra at scales $0.1 \lesssim x \lesssim 1$, while for $M_H = 30$ the spectra at scales $x \lesssim 1$ are basically undistorted. Therefore, the latter seems the most optimal choice.  Note that we could choose even larger values of $M_H$, but then we would be including momenta belonging to the ultraviolet tail. In the limit $M_H \rightarrow \infty$ we recover the divergent zeroth-order-subtracted spectrum at all scales.

Let us now consider the regularized power spectrum of the two-point function. With the PSAR scheme we obtain [c.f.~Eq.~(37) of Ref.~\cite{Ferreiro:2022ibf}]
\be \SprPhi = \frac{H^2 x^3}{4 \pi^2} \left( \pi e^{- \frac{\pi}{2}  \mathfrak{Im}[\nu]} |H_{\nu}^{(1)} (x) |^2 -  \frac{1}{ (m_H^2 + x^2)^{\frac{1}{2}}}  - \frac{1}{(M_H^2 + x^2)^{\frac{3}{2}}}  \right) \ , \label{eq:powsp-dS-alt} \ee
where $M_H \equiv M/H$. We obtain the unregularized spectrum by taking only the first term inside the parentheses, while the second and third terms correspond to the second- and fourth-order subtraction terms, which remove the quadratic and logarithmic divergences respectively. In the massless limit $m \ll H$, the regularized power spectrum  is simply
\be \SprPhi = \frac{H^2}{4 \pi^2} \left( 1 - \frac{x^3}{(M_H^2 + x^2)^{\frac{3}{2}}}\right) \ , \hspace{1cm} [m \ll H] \ , \ee
so it is approximately scale invariant $\SprPhi \simeq H^2 / (4 \pi^2)$ for scales $x \lesssim M_H$. 

\begin{figure*}
    \centering
        \includegraphics[width=0.8\textwidth]{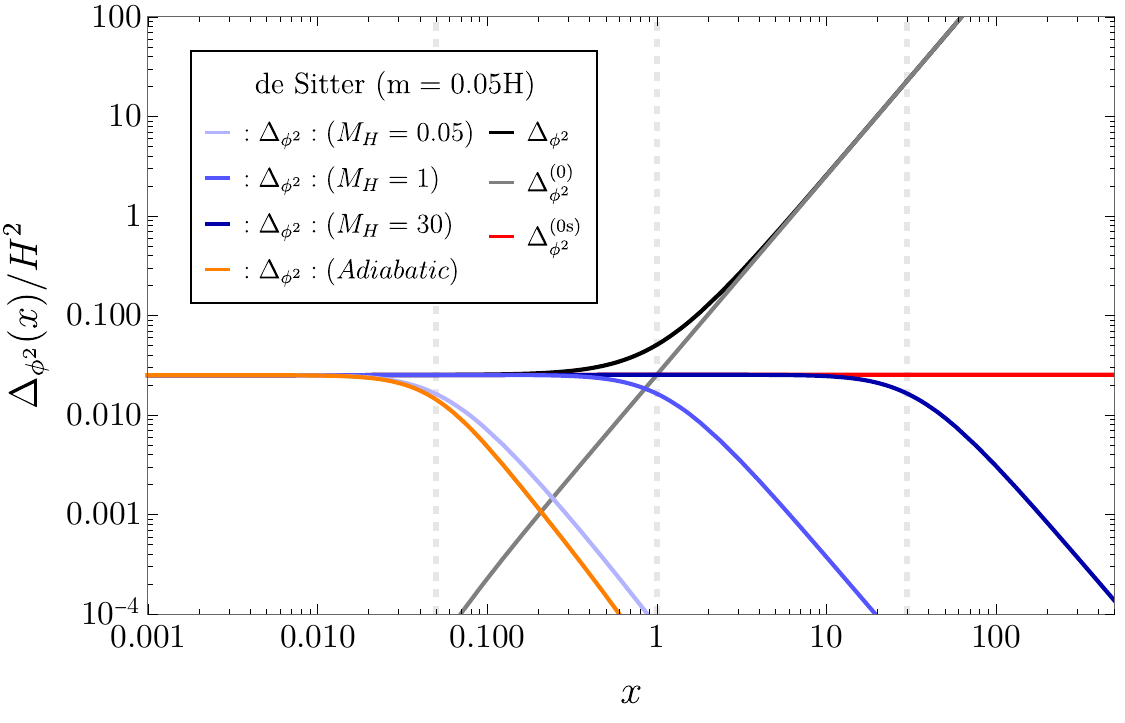}
    \caption{[De Sitter, Sect.~\ref{sec:deSitter}] Power spectra of the two-point function for a scalar field of mass $m=0.05 H$ in de Sitter space with Hubble parameter $H$. The color coding is the same as in Fig.~\ref{fig:PowSpecDs}.} \label{fig:PowSpecDs2}
\end{figure*}

As an example, in Fig.~\ref{fig:PowSpecDs2} we show, for a field with mass $m=0.05 H$, different unregularized and regularized spectra for the two-point function. Note that the zeroth order-regularized power spectrum $\Delta_{\phi^2}^{{\rm (0s)}}$ is approximately scale-invariant [in fact, we have exactly $\Delta_{\phi^2}^{{\rm (0s)}} = H^2 / (4 \pi^2)$ for $m=0$]. As first noticed in \cite{Parker:2007ni}, adiabatic regularization significantly suppresses the amplitude of the power spectrum as super-Hubble scales $m_H \lesssim x (\lesssim 1)$. However, this can be avoided by using the PSAR scheme with $M_H \gtrsim 1$, as observed in the figure.

\subsection{Geometric reheating} \label{sec:geometricreheating}

Let us now consider the regularization of a free scalar field after inflation when it is non-minimally coupled to the scalar curvature. This scenario was originally studied in Ref.~\cite{Bassett:1997az}, which coined the term `geometric reheating' (see also \cite{Tsujikawa:1999jh,Tsujikawa:1999iv}). In this setup, the post-inflationary oscillations of the inflaton give rise to tachyonic oscillations of the Ricci scalar, which generate field instabilities similar to the tachoynic preheating phenomenon studied in \cite{Dufaux:2006ee}. This has been recently studied as a mechanism for dark matter production in the early universe \cite{Markkanen:2015xuw,Fairbairn:2018bsw,Cembranos:2019qlm,Babichev:2020yeo,Lebedev:2022vwf}. In order to fully capture the non-linear dynamics of geometric reheating, one would need to simulate the fields with lattice simulations  \cite{Figueroa:2021iwm}. However, here we restrict ourselves to a linearized analysis, which is a good approximation at early times (when backreaction effects from the excited field onto the background fields can be neglected).

Let us denote as $\Phi$ the inflaton field sourcing the inflationary expansion of the Universe. As an example we consider an inflationary chaotic model with quadratic potential $V(\Phi)=\frac{1}{2} m_\Phi^2 \Phi^2$. We take the inflaton as homogeneous, $\Phi = \Phi (\tau)$, which is a valid approximation during the initial oscillations. The coupled equations of motion for the inflaton and scale factor are
\begin{align}
    & \Phi''+2 H \Phi'+m_\Phi^2a^2\Phi=0 \ , \label{eq:GeRh1} \\
    & H^2(\tau)\equiv \left(\frac{a'}{a} \right)^2=\frac{1}{3 m_p^2}\left(\frac{1}{2} \Phi'^{2}+\frac{1}{2} m_\Phi^2 a^2 \Phi^2\right) \ , \label{eq:GeRh2} 
\end{align}
and the evolution of the Ricci scalar is given by
\be
R (\tau) = 6 \frac{a''}{a^3}=\frac{1}{m_p^2}\left(2 m_\Phi^2 \Phi^2  - \frac{{\Phi'}^2}{a^2}  \right) \label{eq:Ricci} \ .
\ee

We fix our initial conditions at the time when inflation ends $\tau_*$, defined when the first slow-roll parameter becomes $\epsilon_{\mathrm{v}}(\tau_*)= 1$. This condition holds when  $\Phi\left(\tau_*\right) = \sqrt{2} m_p$. In order to obtain the inflaton time derivative at this time $\Phi '\left(\tau_*\right)$, we solve the coupled equations \eqref{eq:GeRh1}-\eqref{eq:GeRh2} with initial conditions deep inside the slow-roll regime, and evolve them until $\tau = \tau_*$. From this we obtain $\Phi^\prime\left(\tau_*\right) \simeq -0.75  m_\Phi m_p$. We also set  $a(\tau_*)=1$.

\begin{figure}
    \centering
        \includegraphics[width=0.59\textwidth]{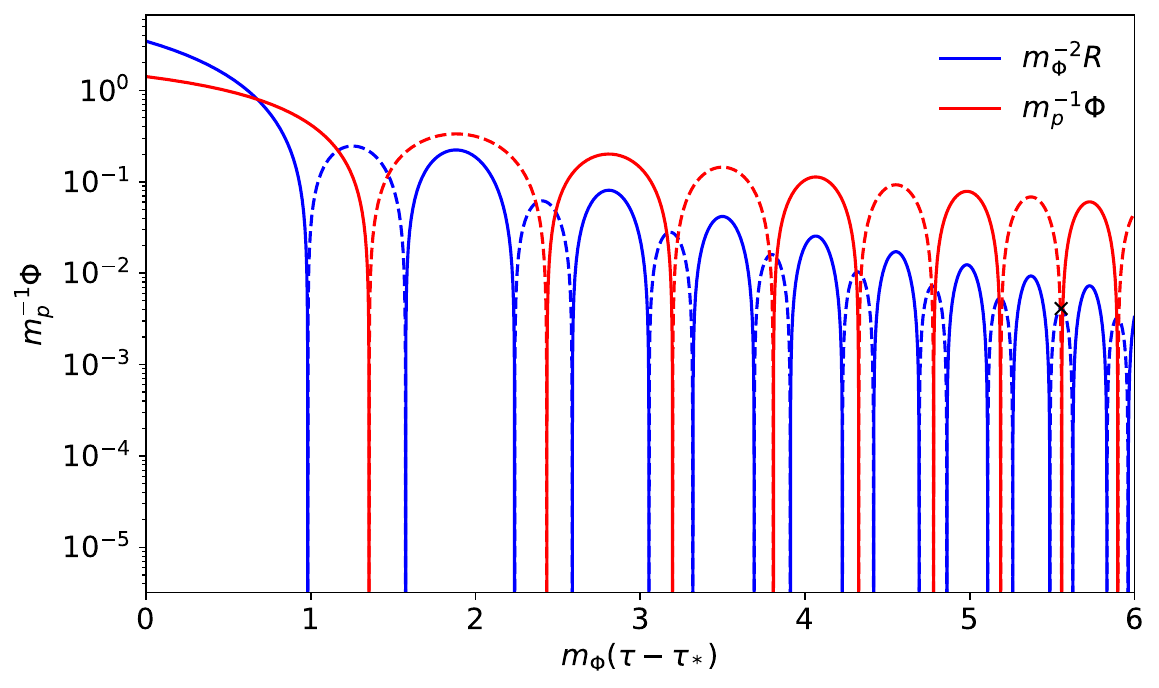}
    \caption{[Geometric reheating, Sect.~\ref{sec:geometricreheating}] Time evolution of the inflaton amplitude $\Phi/ m_p$ (red) and the Ricci scalar $R /m_{\Phi}^2$ after quadratic chaotic inflation. The solid/dashed pattern in each line represents positive/negative values.} \vspace*{-0.2cm} \label{fig:R}
\end{figure}

In Fig.~\ref{fig:R} we show the evolution of $\Phi = \Phi (\tau)$ and $R = R (\tau)$ after the end of inflation. The Ricci scalar oscillates between positive and negative values: it attains its local (positive) maxima when $\Phi' = 0$, while it attains its local (negative) minima when $\Phi \simeq 0$. If the quantized scalar $\chi$ has a non-minimal coupling to curvature $\xi > 1/6$, its effective mass becomes negative during part of each inflaton oscillation. This triggers a tachyonic amplification of the field modes. More specifically, we can write the field mode equation \eqref{eq:fieldmodeeq} as a \textit{Mathieu} equation, which has a well-studied structure of resonance bands for which the field modes grow exponentially \cite{Bassett:1997az}. As expected, the resonance is stronger for larger values of $\xi$.

In order to track the amplification of the quantized scalar field modes after inflation, we numerically solve the field mode equation \eqref{eq:fieldmodeeq} for the Ricci scalar \eqref{eq:Ricci} in the light field limit $m^2 \ll (\xi - \frac{1}{6}) \langle R \rangle_{T_{\Phi}}$  (where $\langle \dots \rangle_{T_{\Phi}}$ denotes an oscillation average), for different values of $\xi \in [1/3, 10]$. More specifically, we solve the field mode equation for a set of $10^3$ logarithmically spaced discrete momenta within the interval $\kappa \equiv k / m_{\phi} \in [10^{-2}, 50]$, which allows us to track the evolution of the different power spectra. The equations are solved with the {\tt scipy.solve\_ivp} method available in the {\tt SciPy} package of {\tt Python}.

The initial conditions for the field modes must be carefully chosen so that they have the correct behavior in the ultraviolet limit (if not, the subtraction terms will not cancel the ultraviolet divergences). We can achieve this by means of the $\mu$-adiabatic expansion introduced in \eqref{WKBansatz2}. More specifically, we set as initial conditions the following fourth-order adiabatic vacuum state,
\be 
\chi_k(\tau_*;\mu_0)=\frac{1}{\sqrt{2\overline{W_k}^{(0-4)}(\tau_*;\mu_0)}}\ , \hspace{0.4cm}
\chi_k^\prime(\tau_*;\mu_0)=-i\sqrt{\frac{\overline{W_k}^{(0-4)}(\tau_*;\mu_0)}{2}}-\frac{\overline{W_k}^{'(0-4)}(\tau_*;\mu_0)}{\left(2\overline{W_k}^{(0-4)}(\tau_*;\mu_0) \right)^{3/2}} \ , 
\ee
where we set the arbitrary mass scale $\mu_0$ to the effective mass at the end of inflation $\mu_0 \equiv (\xi - \frac{1}{6}) a^2 (\tau_*) R(\tau_*)$. One can show that $\overline{W_k}^{(2)} (\tau_*;\mu_0) / \overline{W_k}^{(0)}  (\tau_*;\mu_0) < 1.42$ and $\overline{W_k}^{(4)}  (\tau_*;\mu_0)/ \overline{W_k}^{(0)}  (\tau_*;\mu_0) < 1.28$ for all considered momenta and coupling constants, so higher orders in the expansion only provide small corrections to the zeroth-order initial condition.

\begin{figure*}
    \centering
    \includegraphics[width=0.85\textwidth]{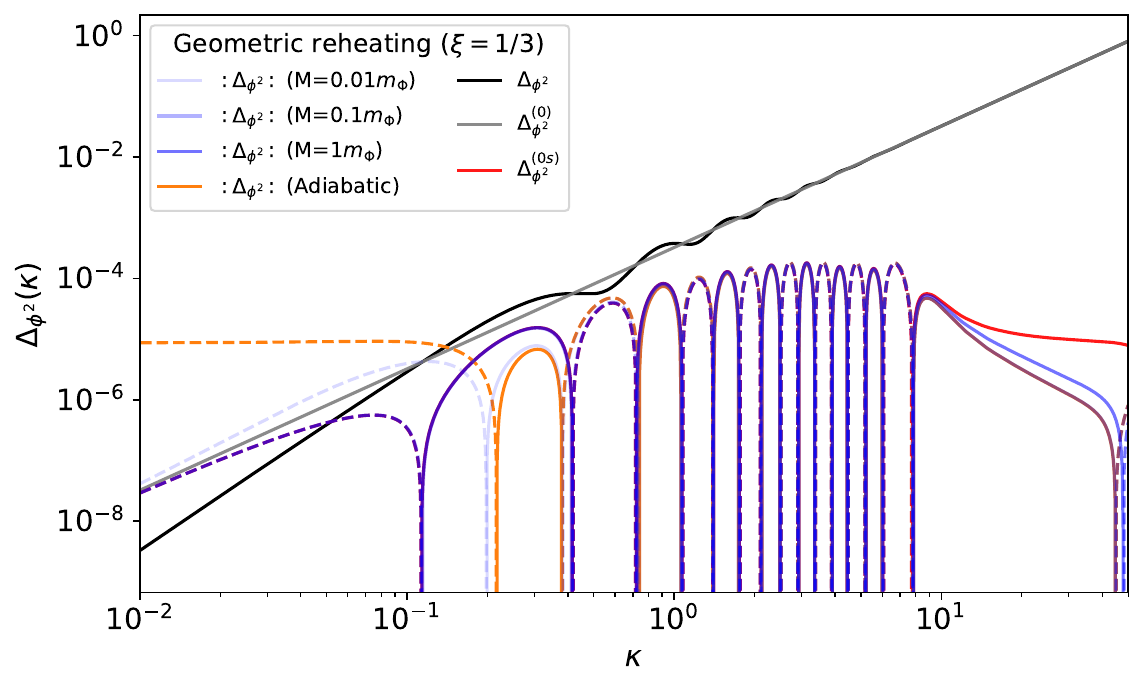} 
    \includegraphics[width=0.85\textwidth]{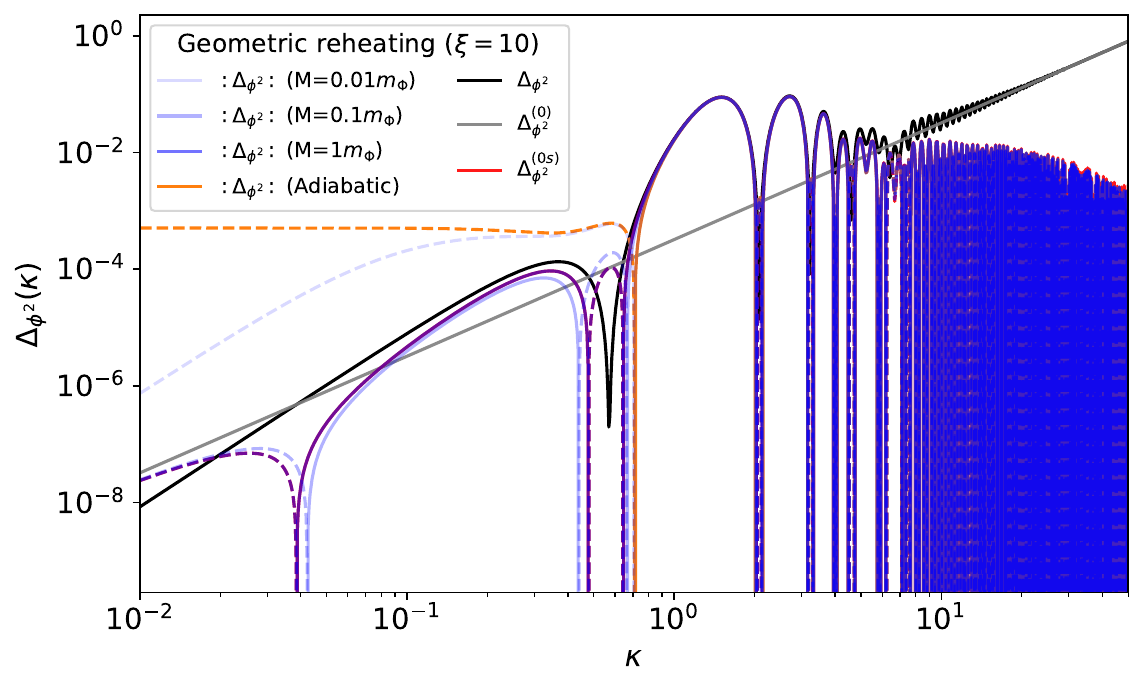} 
    \caption{[Geometric reheating, Sect.~\ref{sec:geometricreheating}] Power spectra of the two-point function for a scalar field in the light field limit $m^2 \ll (\xi - 1/6)\langle R\rangle_{T_{\Phi}}$ with $\xi=1/3$ (top panel) and $\xi=10$ (bottom panel), as a function of $\kappa \equiv k /m_{\Phi}$. The spectra are computed when the Ricci scalar achieves its eighth minimum. In each panel we show: the unregularized spectra (black line), the zeroth order subtraction term (gray), the zeroth-order-subtracted one \eqref{eq:MinkSubsPW} (red), the spectra regularized with the adiabatic scheme (orange), and the spectra regularized with our proposed PSAR scheme for different choices of $M$ (blue lines). The dashed pattern indicates negative values of the quantity.}  \label{fig:PowSpecGr}
\end{figure*}

In Fig.~\ref{fig:PowSpecGr} we depict different unregularized and regularized power spectra for the two-point function of a scalar field in the light field limit $m^2 \ll (\xi - \frac{1}{6}) \langle R \rangle_{T_{\Phi}} $ (in practice we impose exactly $m=0$ in the numerical solver). We show the cases $\xi=1/3$ (top panel) and $\xi = 10$ (bottom panel), and depict the spectra when the Ricci scalar attains a local minimum for the eighth time (this corresponds approximately to four inflaton oscillations). In the case $\xi = 10$, a structure of several peaks gets imprinted  in the zeroth-order-subtracted spectrum $\Delta_{\phi^2}^{\rm (0s)} \equiv \Delta_{\phi^2} - \Delta_{\phi^2}^{(0)}$ at momenta scales $\kappa \equiv k / m_{\Phi} \lesssim 10$, due to the resonance bands of the field mode equation. In these peaks we have $\Delta_{\phi^2} / \Delta_{\phi^2}^{\rm (0)} \gtrsim 1 $. For $\xi = 1/3$, a structure of peaks also gets formed, but in this case the resonance is much weaker and we have $\Delta_{\phi^2} / \Delta_{\phi^2}^{\rm (0)} \ll 1 $ even in the peaks.

In both panels of Fig.~\ref{fig:PowSpecGr} we depict the spectra regularized with the standard adiabatic method (orange) and the PSAR method for values $M = 0.01 m_{\Phi}, 0.1 m_{\Phi}, m_{\Phi}$ (blue lines). As expected, both methods get rid of the ultraviolet divergences. However, as explained in Sect.~\ref{sec:Regularization}, we require our regularization schemes to obey $\SprPhi \simeq \Delta_{\phi^2}^{{\rm (0s)}}$ at all amplified infrared modes (i.e.~at the peaks of the spectra). We can observe that the standard adiabatic scheme completely distorts the amplitude of the power spectrum at momenta $\kappa \lesssim 0.5$, and in fact, the most infrared band for $\xi = 10$ completely vanishes after regularization. However, if we use the PSAR method, the infrared distortions introduced by regularization are significantly less important. In fact, for the choice $M \gtrsim m_{\Phi}$ we have almost exactly $\SprPhi \simeq \Delta_{\phi^2}^{{\rm (0s)}}$ at all momenta $\kappa \lesssim 10$.

\begin{figure*}
    \centering
    \includegraphics[width=0.88\textwidth]{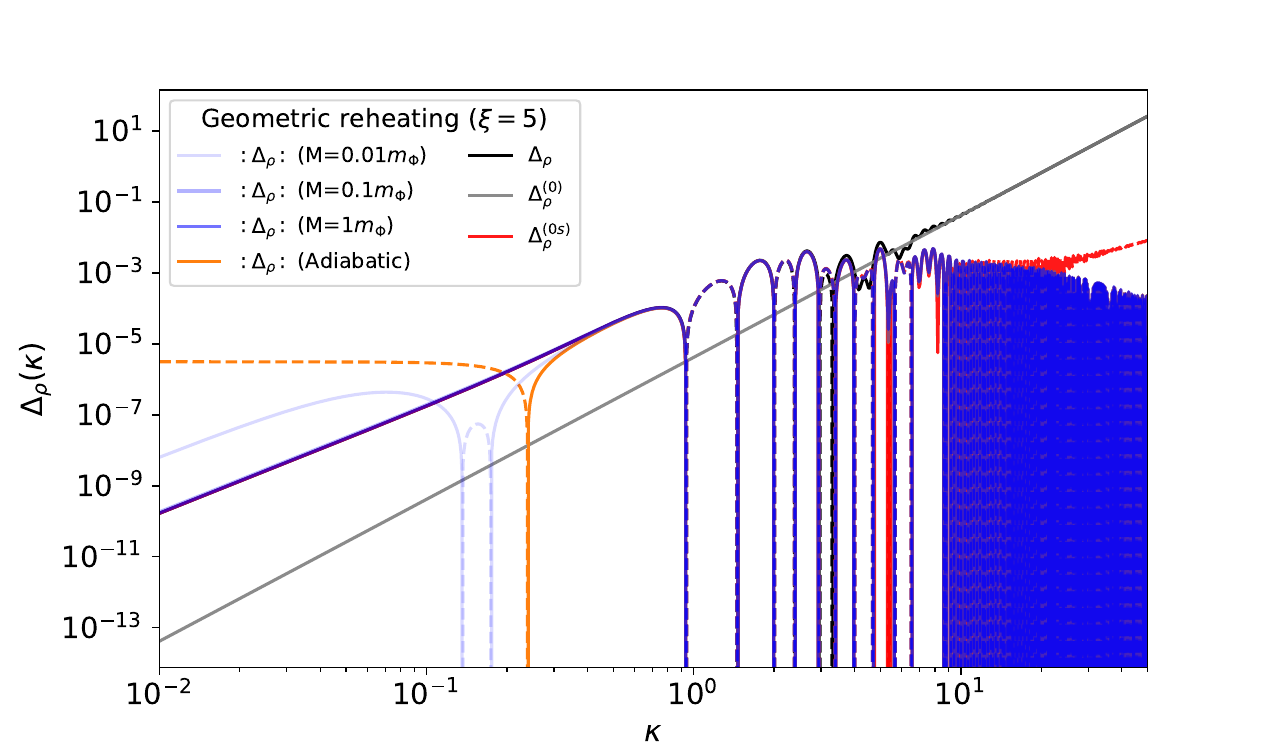} 
    \includegraphics[width=0.88\textwidth]{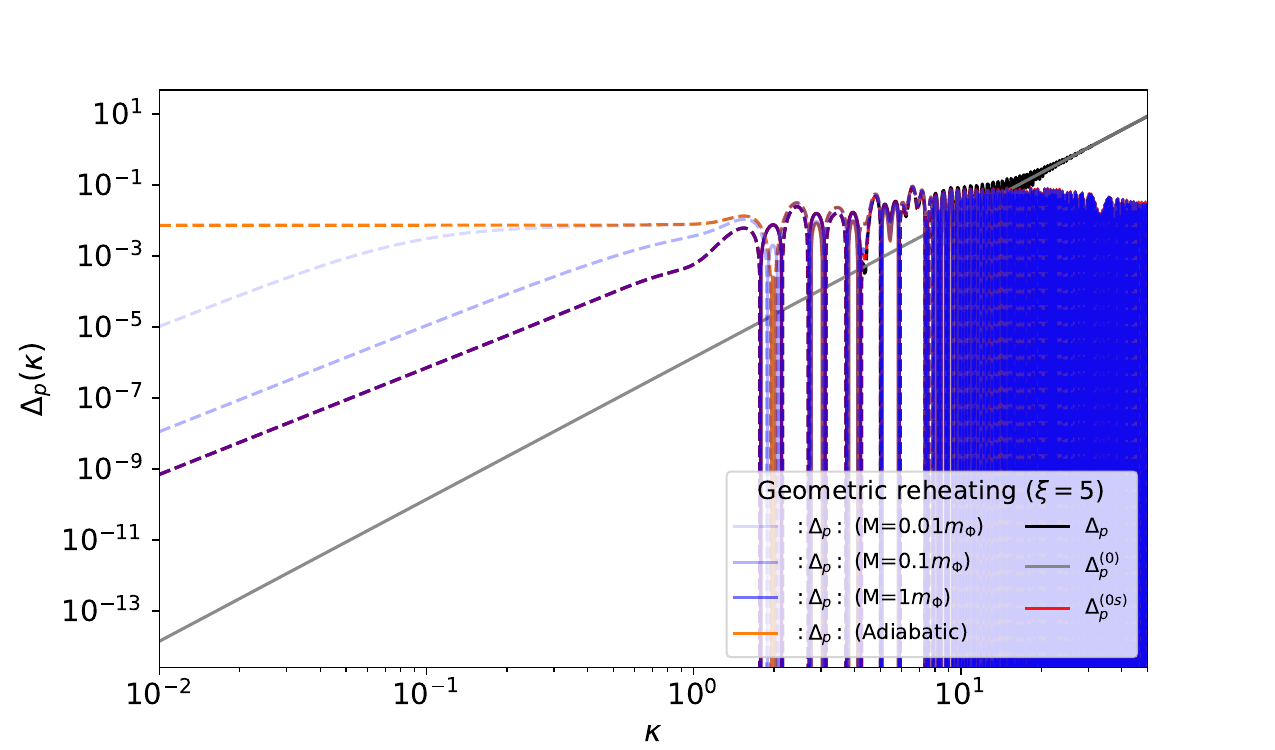} 
    \caption{[Geometric reheating, Sect.~\ref{sec:geometricreheating}] Power spectra of the energy and pressure densities (top and bottom panels respectively) for a scalar field in the light field limit $m^2 \ll (\xi - 1/6)\langle R\rangle_{T_{\Phi}}$ with $\xi=5$, as a function of $\kappa \equiv k / m_{\Phi}$. The spectra are computed at the time where the Ricci scalar achieves its eighth minimum. We use the same color coding as in Fig.~\ref{fig:PowSpecGr}.}  \label{fig:PowSpecTxi5}
\end{figure*}

Finally, we show in Fig.~\ref{fig:PowSpecTxi5} several (unregularized and regularized) power spectra for the energy and pressure densities (top and bottom panels respectively). We consider again a light scalar field, and we have fixed $\xi=5$ in both cases.  The above analysis on the effects of the regularization scheme at infrared scales also applies for these quantities. We can observe that the standard adiabatic method distorts the spectrum at scales $\kappa \lesssim 1$, and as for the two-point function, the most infrared peak present in $\Delta_{c}$ completely vanishes in the adiabatic-regularized spectra. However, if we use the PSAR method, we have almost exactly $\SprRho \simeq \Delta_{\rho}^{{\rm (0s)}}$ and  $\SprPre \simeq \Delta_{p}^{{\rm (0s)}}$  at all momenta $\kappa \lesssim 10$.

\section{Summary and discussion} \label{sec:Summary}

During the last decades, the development of a self-consistent renormalization program for free quantum fields in curved spacetimes compatible with covariance and locality has been successfully achieved for scalars, fermions and vector fields. The different designed methods produce unique stress-energy tensors and two-point functions up to arbitrary finite geometrical terms as expressed in Eq.~\eqref{arbT}. The ambiguous subtracted divergent terms can then be reabsorbed in the coupling constants in the Einstein's equations, which can be measured experimentally. Although a mathematical framework for this method has been successfully established, less effort has been made in the development of an efficient methodology to do actual computations, both analytically and numerically. This work has aimed to fill in this gap in the context of cosmological spacetimes.

Even in the simplest case of free scalar fields, the expectation value of the stress-energy tensor (or the two-point function at coincident spacetime points) has ultraviolet divergences that cannot be removed through a normal ordering procedure as in Minkowski spacetime.  A well-established regularization method when working in cosmological spacetimes is adiabatic regularization, which allows to identify the divergent terms of any quantity by means of an adiabatic expansion of the field modes. The method is compatible with locality and covariance, and is equivalent to regularization methods for generic curved backgrounds when particularized to FLRW spacetimes \cite{delRio:2014bpa, Beltran-Palau:2021xrk}. It is also feasible to use in numerical computations, see e.g.~\cite{Anderson:1987yt,Habib:1999cs,Zago:2018huk}. However, the adiabatic subtraction terms for a given quantity can substantially modify the amplitude of its power spectrum at the momenta scales amplified by the non-adiabatic expansion, which we expect to be classical and then not affected by the quantum subtraction terms. We have denoted these modifications as `infrared distortions' (in contrast with the ultraviolet tail, which is the one we expect to remove through regularization).

The aim of this work has been to build a new regularization method that both includes the ambiguities allowed by Eq.~\eqref{arbT} and allows to build observables without infrared distortions. The result has been the set of subtraction terms \eqref{eq:subs-rho}-\eqref{eq:subs-pres} (for the stress-energy tensor) and \eqref{eq:two-pf-sub} (for the two-point function), derived in Sect.~\ref{sec:Regularization}. The method is equivalent to the normal-ordering procedure in Minkowski spacetime, produces a conserved stress-energy tensor, and allows to construct regularized observables without infrared distortions. We have named our proposed method \textit{physical scale adiabatic regularization} (PSAR), in order to differentiate it with respect to the standard scheme. Our method includes a set of arbitrary mass scales ($M$, $\Mma$, $\Mmb$), allowed by Eq.~\eqref{arbT}, which can tame the infrared distortions of any observable if set large enough. Moreover, in Sect.~\ref{sec:RenormCond} we have interpreted the subtraction terms as redefinitions of the coupling constants set by an appropriately chosen renormalization condition. Finally, in Sect.~\ref{sec:applications} we have illustrated our method in two examples of cosmological interest: a scalar field propagating in a de Sitter space, and a scalar field being excited after inflation through a process of geometric reheating.

Let us now comment about possible extensions and applications of our work. First, in this work we have developed the PSAR method for scalar fields, but it would be interesting to extend our proposed method to other field species such as fermions. The standard adiabatic expansion for spin-1/2 fields was introduced in Ref.~\cite{Landete:2013axa, Landete:2013lpa, delRio:2014cha, BarberoG:2018oqi} and, unlike scalars, it is not based on a WKB template. However, it is not clear to us if a method to minimize infrared distortions is possible (or even physically meaningful) in this case: their occupation number is always $n_k < 1$ due to Pauli blocking, so there is not such a clear hierarchy between infrared and ultraviolet modes like there is for scalars. Similarly, it would be interesting to generalize our method to quantized fields coupled to homogeneous time-dependent backgrounds like in preheating scenarios, including both scalar preheating \cite{Kofman:1994rk,Kofman:1997yn,Greene:1997fu} and fermionic preheating \cite{Greene:1998nh,Giudice:1999fb,Greene:2000ew,Peloso:2000hy}. Adiabatic expansions for these scenarios have been developed for scalars \cite{Anderson:2008dg} and fermions \cite{delRio:2017iib}, which could be a nice starting point to build a PSAR method for these setups. Our method could also be potentially extended to the case of quantum fields with interactions to classical electric fields 
\cite{Ferreiro:2018qdi,Ferreiro:2018qzr,Beltran-Palau:2020hdr}. 

Finally, it would be interesting to apply our regularization method to the inflationary correlation functions. Their regularity at infrared scales has been extensively studied in the literature, see e.g.~\cite{Tanaka:2013caa,Hu:2018nxy,Hu:2020luk} and references therein, but ultraviolet effects are typically neglected. Our results suggest that this might be a reasonable assumption. However, note that here we have restricted ourselves to free scalar fields evolving in fixed FLRW metrics, while for such study we would need to consider both field interactions and metric perturbations.

\acknowledgments
We thank Jose Navarro-Salas for very useful remarks about the manuscript. F.T. is supported by a \textit{Mar\'ia Zambrano} fellowship (UV-ZA21-034) from the Spanish Ministry of Universities and grant PID2020‐116567GB‐C21 of the Spanish Ministry of Science. \vspace*{-0.3cm}
  
  \bibliography{References.bib}

\end{document}